\documentclass[12pt]{article}
\usepackage{jheppub}
\usepackage{mymacros}

\title{A teleportation protocol in Schwarzschild-de Sitter space}
\author[a]{Sergio E. Aguilar-Gutierrez,}
\author[b]{Ricardo Espíndola,}
\author[c,d]{and Edward K. Morvan-Benhaim}
\affiliation[a]{Institute for Theoretical Physics, KU Leuven, Celestijnenlaan 200D, B-3001 Leuven, Belgium}
\affiliation[b]{Institute for Advanced Study, Tsinghua University, Beijing 100084, China}
\affiliation[c]{Institute of Physics, University of Amsterdam, Science Park 904, PO Box 94485, 1090 GL Amsterdam, The Netherlands}
\affiliation[d]{Delta Institute for Theoretical Physics, Science Park 904, PO Box 94485, 1090 GL Amsterdam, The Netherlands}

\emailAdd{sergio.ernesto.aguilar@gmail.com, ricardo.esro1@gmail.com, edward.morvan@gmail.com}
\abstract{We propose a new information transfer protocol for de Sitter space, using black holes as energy reservoirs. We consider antipodal observers in pure de Sitter space in the Bunch-Davis state. They can store Hawking modes from the cosmological horizon in a box. Alternatively, due to thermal fluctuations in de Sitter space, {black holes formed through a} pair-creation {process can be used} as energy reservoirs. We focus on the Nariai {black hole} case, which corresponds to an equilibrium state. Once {the black hole is produced}, energy pulses can be released into its interior, opening a traversable wormhole. We provide bounds for the amount of information that can be transferred. Specializing in (1+1)-dimensions, we explore how the teleportation protocol leads to an explicit geometric description of the information transmitted through an island region. The protocol uncovers quantum information aspects of de Sitter space, {
independently of any} particular realization of de Sitter space holography.}

\begin{document}

\maketitle

\section{Introduction}
In recent years, traversable wormholes (see \cite{Kundu:2021nwp} for a recent review) have provided key insights into how quantum information builds spacetime in the context of the ER=EPR conjecture \cite{Maldacena:2013xja}. To generate a traversable wormhole in any crunching spacetime region, there needs to be a violation of the averaged null energy condition (ANEC) \cite{Freivogel:2019lej}. This condition states that
\begin{equation}
    \int_{\Gamma} T_{\mu\nu}k^\mu k^\nu \rmd \lambda\geq0~,
\end{equation}
where $T_{\mu\nu}$ is the matter stress-energy tensor along complete achronal null geodesics, $\Gamma$, which is described by the tangential vectors $k^\mu$ and the affine parameter $\lambda$. This is essentially a statement about causality in general relativity \cite{Morris:1988tu,Friedman:1993ty,Wall:2009wi}, which can be violated by quantum matter in curved spacetime \cite{Visser:1996iw,Visser:1995cc}. Gao, Jafferis, and Wall (GJW) first proposed a protocol that generates traversable wormholes in BTZ black holes \cite{Gao:2016bin}. Several improvements to this protocol have been made in the last few years. Some examples include {the} stabilization of the wormhole with different matter content \cite{Maldacena:2017axo,Maldacena:2018lmt,Maldacena:2018gjk,Maldacena:2020sxe, Bintanja:2021xfs}, and {applications to probe} typical black hole microstates \cite{deBoer:2018ibj,DeBoer:2019yoe}. Although the previous findings rely on the AdS/CFT correspondence, there has been much interest in extending holography to arbitrary types of spacetimes, perhaps even those pertinent for early and late-time cosmology \cite{Hartman:2020khs,Aguilar-Gutierrez:2021bns,Bousso:2022gth,Espindola:2022fqb,Ben-Dayan:2022nmb}. In particular, there is a recent proposal for information exchange in pure dS space \cite{Aalsma:2021kle} which {implements} a protocol for generating traversable wormholes in de Sitter (dS) space according to a pair of antipodal static patch observers, using the fact that positive energy perturbations elongate the dS Penrose diagram, as shown by Gao and Wald \cite{Gao:2000ga}. The main technical difficulty in adapting the GJW protocol to dS space is it lacks a well-developed holographic dual theory. {Thus, one has to rely} solely on a bulk perspective, resulting in two major difficulties: Assuring the synchronicity of the clocks between the two static observers wanting to exchange information, {and} controlling the amount of energy required to open up the wormhole.

\paragraph{Purpose of the work} We propose a modification of the GJW protocol for black holes in dS space and derive bounds on the amount of information that can be extracted from {the} protocol.
We first {use a} protocol described in \cite{Aalsma:2021kle} {to allow} two causally separated observers to communicate with each other. Although each information exchange {between the observers} would only {allow an amount} order $\mathcal{O}(1)$ {of bits,} and require a dS scrambling time $t_S=\ell\log{S_{dS}}$ {for the message} to arrive, the two observers could agree on a binary language akin to Morse code to decode it. In return, this classical communication channel (as in the BTZ case \cite{Gao:2016bin}) {makes it possible} to synchronize their clocks and schedule future events.
Secondly, as pointed out in \cite{Aalsma:2021kle}, more energy needs to be gathered from the cosmological horizon to increase the {amount of} information exchange between the two static patches. This can be achieved by placing a one-way transparent, or semi-mirrored (i.e. imposing Dirichlet boundary conditions on one side and transparent boundary conditions on the other one) Dyson sphere (DS) \cite{Dyson:1960xib} around the center of the static patch $r=0$, effectively producing an energy reservoir. To guarantee the largest {information} exchange, we consider a DS of radius $r_{\rm DS}=r_N$, corresponding to the largest black hole radius that can fit into a patch of dS ({i.e. a }Nariai {black hole}). After a collecting time $\tilde{\mathcal{T}}_N\propto r_N^3$, both observers will agree to let the radiation escape from the DS. On the one hand, the released energy will act as (a series of) positive shock waves emanating from the DS, {and }result in the formation of a cosmic wormhole \cite{Aalsma:2021kle}. On the other hand, the \textit{lack} of energy bouncing back towards the black hole effectively acts as (a series of) \textit{negative} shocks due to energy conservation, which opens up a wormhole through the black hole region \cite{Freivogel:2019whb}.

Alternatively, when waiting long enough, a pair of black holes can spontaneously appear as a consequence of the thermal bath filling empty dS spacetime\footnote{See \cite{Ginsparg:1982rs} for the original description of the semiclassical instability of dS space, and \cite{Bousso:1996au} for the pair creation rate
from the no boundary proposal. See also \cite{Draper:2022xzl,Morvan:2022ybp,Morvan:2022aon} for recent discussions about a constrained instanton interpretation {of} this process.}.\\

As {previously} mentioned, one of the advantages of our procedure is that it does not require a particular realization of dS space holography\footnote{We refer the reader to \cite{Witten:2001kn,Strominger:2001pn,Maldacena:2002vr} for pioneering work in this area, and \cite{Galante:2023uyf} for a recent review.}. Moreover, although we explicitly perform {the} protocol in (3+1)-dimensions to describe how the observers can trap the radiation, a generalization to arbitrary dimensions should be possible, given the generality of the arguments. We will provide bounds on the amount of information that can be exchanged by the observers based on the sizes of the respective wormhole throats.\\
\\
After obtaining the information bound based on the classical gravitational backreaction, we consider the effects arising from the presence of (entangled) matter fields across the two static patches, leading to the formation of (at least) one island in the black hole region. See \cite{Penington:2019npb,Almheiri:2019psf} for the original formulation of the island rule, and \cite{Almheiri:2019hni} for a pedagogical review. According to this rule, the von Neumann entropy of a weakly gravitating region $\mathbf{R}$ is given by the minimum value among all extrema of the generalized entropy of a region $\mathbf{R}\cup\mathbf{I}$. {In other words}
\begin{equation}\label{eq:SR vN}
    S(\mathbf{R}) = \text{min~} \text{ext}_{\mathbf{I}}S_{\rm gen}(\mathbf{R}\cup\mathbf{I})~,
\end{equation}
where $\mathbf{I}$ is called the island region found by extremizing the functional
\begin{equation} \label{eq:Sgen}
   S_{\rm gen}(\mathbf{R}\cup\mathbf{I})=\frac{A(\partial \mathbf{I})}{4G_N}+S_{\rm matter}(\mathbf{R}\cup\mathbf{I})~.
\end{equation}
{In this equation, }$S_{\rm matter}(\mathbf{R}\cup\mathbf{I})$ denotes the von Neumann entropy of the bulk matter fields, and $A(\partial \mathbf{I})$ is the codimension-2 area functional evaluated at the endpoints of the {region} $\mathbf{I}$.

The observer on the region $\mathbf{R}$ can perform a series of decoding channels from the Hawking radiation, such as the Petz map \cite{Chen:2019gbt,Penington:2019kki,Zhao:2020wgp,Bak:2021qbo,Vardian:2023fce,Bahiru:2022ukn}, reconstruct the entanglement wedge that includes the island region. 
In this case, information is sent to the island region and later collected {by} low-energy observers.\footnote{Related problems on information transmission through wormholes and islands have been addressed in the DS/dS context \cite{Geng:2020kxh,Geng:2021wcq}.} Although the island rule is less developed in the context of dS space, there are indications that islands, seen as disconnected entanglement wedges, will appear in general types of backgrounds \cite{Bousso:2022hlz,Bousso:2023sya}.

To make an explicit realization of the above {protocol}, we employ the s-wave dimensional reduction of a near-Nariai black hole background, called the full-reduction dS Jackiw–Teitelboim gravity (JT) gravity model. See \cite{Chen:2020tes,Piao:2023vgm,Sybesma:2020fxg,Goswami:2022ylc,Aguilar-Gutierrez:2021bns,Aalsma:2021bit,Levine:2022wos,Balasubramanian:2020xqf,Kames-King:2021etp,Hartman:2020khs,Baek:2022ozg,Aalsma:2022swk,Teresi:2021qff,Jiang:2024xnd} for different developments on the island rule in dS JT gravity. The antipodal static patch observers will prepare the shockwaves from the thermal bath. We then locate a pair of detectors at $\mathcal{I}^+$ of the inflating regions, representing meta observers, to collect Hawking modes coming from the black hole, as shown in Figure \ref{fig:Setup_Islands_detectors}. In this situation, the effect of the shockwave is to shift the location of the island with respect to the one without the backreaction. The antipodal observers can transmit information about the island region directly to the {detectors (i.e. the} metaobservers).

Our \emph{goal} is then to derive bounds on the information transfer in near-Nariai black hole backgrounds, independent of dS holographic assumptions, {using the GJW protocol,} as well as {from} the island rule in the s-wave reduction of the near-Nariai geometry with CFT$_2$ bulk matter fields.

\paragraph{Outline} The structure of the manuscript is as follows. In Sec. \ref{sec:Set up}, we review some of the basics {on} state preparation and SdS black holes to carry out our protocol. We will then focus mostly on (3+1)-spacetime dimensions to make concrete statements on how to trap the radiation. We find bounds on the amount of information that can be transferred between static patch observers. Sec. \ref{Sec:2d} is devoted to the (1+1)-dimensional formulation of the protocol from dS JT gravity where we add metaobservers in the inflationary patch of the near-Nariai spacetime equipped with detectors. We show that the traversable wormhole protocol can probe the island region according to the metaobservers. In Sec. \ref{Sec:Discussion}, we include a summary of our proposal and we conclude with some future directions. For the convenience of the reader, App. \ref{App:HaydenPreskill} reviews the Hayden-Preskill (HP) protocol to perform information recovery, which {we apply in} our protocol. Meanwhile, App. \ref{App:JT from SdS} contains a review of the s-wave reduction of SdS black holes in higher dimensions to generate the full-reduction dS JT gravity theory. Lastly, App. \ref{App:Shift SdS} contains {a} discussion about the small shift approximation in the shockwave analysis in the weakly coupled gravity regime.

\section{General protocol}\label{sec:Set up}
\subsection{State preparation}
\begin{figure}[t!]
    \centering
    \begin{subfigure}[c]{0.49\textwidth}    \includegraphics[width=\textwidth]{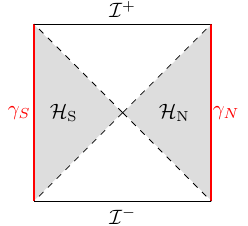}\end{subfigure}\hfill\begin{subfigure}[c]{0.49\textwidth}    \includegraphics[width=\textwidth]{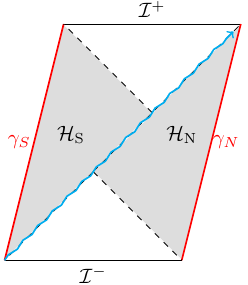}\end{subfigure}
    \caption{(a) dS space as seen by a pair of antipodal observers following a worldline $\gamma_N$ and $\gamma_S$ (in red) for the Bunch-Davis preparation of state described as a thermofield double state, with respective gravitational Hilbert spaces $\mathcal{H}_{N/S}$. (b) The spacetime gets elongated after the insertion of a positive energy shockwave (cyan wavily line) \cite{Gao:2000ga}. The dashed black line indicates the location of the cosmological horizon.}
    \label{fig:dS plain}
\end{figure}

Our starting configuration follows the considerations in \cite{Aalsma:2021kle,Chandrasekaran:2022cip,Witten:2023qsv} for constructing the Bunch-Davis state of pure dS space. We begin studying a single static patch region. Take for example the south {wedge}, $S$. In the perturbative quantum gravity regime, we can define a Hilbert space with energy eigenstates, $\ket{E_i}_S\in\mathcal{H}_S$ with respect to a one-sided modular Hamiltonian $H_S$. One can also define a natural thermal vacuum state, known as the Bunch-Davis state by introducing a decoupled antipodal static patch region. Let us denote it by {the} north, $N$, with energy eigenstates $\ket{E_i}_N\in\mathcal{H}_N$ for the Hamiltonian $H_N$. We may generate the cyclic, separating vacuum of the theory by {means of the} thermofield double construction of the maximally entangled state with respect to $S$ and $N$ \cite{Chandrasekaran:2022cip},
\begin{equation}\label{eq: BD state}
    \ket{\Psi_{\rm BD}}=\frac{1}{\sqrt{Z}}\sum_i\exp\qty(-\frac{1}{2}\beta E_i)\ket{E_i}_S\otimes\ket{E_i}^*_N\in \mathcal{H}_S\times\mathcal{H}_N~.
\end{equation}
{The symbol} $*$ denotes complex conjugation, and $\beta$ refers to the inverse temperature of both patches. The Bunch-Davis state is the vacuum of the time automorphism generator, $H_{\rm grav}=H_N-H_S$, where the minus sign indicates the difference in time orientations between antipodal patches; and {importantly,}  $H_{\rm grav}\ket{\Psi_{\rm BD}}=0$. This construction has been recently discussed in the context of the algebra of observables for dS space in \cite{Chandrasekaran:2022cip} in the presence of matter and weakly coupled gravity (see \cite{Gomez:2022eui,Seo:2022pqj,Gomez:2023wrq,Gomez:2023upk,Jensen:2023yxy,Gomez:2023tkr,Aguilar-Gutierrez:2023odp} for related developments). The initial configuration is illustrated in Fig. \ref{fig:dS plain}.
The pair of antipodal observers are then prepared in a maximally entangled {state} and {they }possess a classical communication channel, so {we will assume that their} internal clocks are synchronized as we start the protocol.

\subsection{Black holes in dS space}\label{Sec:BHs in dS}
The main motivation in this work is to incorporate the appearance of black holes into {an} information exchange protocol in dS space. We will allow the observers to construct an energy reservoir from the Hawking radiation coming out of the cosmological horizon in Fig. \ref{fig:dS plain}. The reservoir can then collapse into {an} SdS black hole connecting the antipodal observers. We will specify the exact conditions for this to happen in Sec. \ref{sec:preparation shockwaves}, after introducing some general notions in this section. Alternatively, we may also consider the semiclassical instability process of dS space {that leads to} pair production of SdS black holes \cite{Ginsparg:1982rs}.

SdS black holes in $(d+1)$-dimensions are spherically symmetric vacuum solutions to Einstein equations with a positive cosmological constant $\Lambda>0$. These configurations can be described with the following metric in static patch coordinates
 \begin{align}
  \rmd s^2 &= -f(r) \rmd t^2 + \frac{\rmd r^2}{f(r)} + r^2 \rmd \Omega_{d-1}^2 \, , \quad \text{with} \label{metric1} \\ 
  f (r) &= 1 - \frac{r^2}{\ell^2} - \frac{16 \pi G_N M  }{(d-1) \Omega_{d-1} r^{d-2}}\,. \label{eq:blackeningfactor}
  \end{align}
Here, $M$ is the black hole mass parameter, $\ell^2=d(d-1) /(2 \Lambda)$ is the dS curvature radius, $G_N$ is Newton's constant  and $\Omega_{d-1}=2\pi^{d/2}/ \Gamma(d/2)$ is the volume of a unit ($d-1$)-sphere. In (\ref{metric1}), we have performed a gauge fixing choice where the static patch time of the $N$ and $S$ pole observers are described as $t=t_{S}=t_{N}$. On the other hand, the parameter $M$ cannot be arbitrarily large, as this can result in complex roots for the blackening factor $f(r)$, which would correspond to a naked singularity. Such structures are forbidden in nature according to the cosmic censorship conjecture; see \cite{Ong:2020xwv} for a review. Therefore, we consider a range for the mass of SdS black holes $M\in [0,\,M_{\rm N}]$. For $M=0$, one recovers empty dS space, with a cosmological horizon located at $r=\ell$.  The maximal mass $M_{\rm N}=\frac{d-1}{d}\frac{\Omega_{d-1}}{8\pi G_N}r_{\rm N}^{d-2}$ is referred to as the Nariai mass. {This limit} corresponds to the extreme case where the radii of the black hole and cosmological horizon {coincide}, $r_{\rm b}=r_{\rm c}=r_{\rm N}=\ell\sqrt{\frac{d-2}{d}}$, {which is called} the Nariai radius \cite{Nariai,Ginsparg:1982rs}.\footnote{The reader might check \cite{Morvan:2022ybp} for further details on the quasilocal thermodynamics of these solutions.} For our purposes, we describe the temperature, $T$, and the thermodynamic entropy, $S$, of the cosmological, $\rm c$, and black hole, $\rm b$, horizons by
\begin{equation}\label{eq:thermo SdS}
    T_{\rm b,\,c}=\frac{\kappa_{\rm b,\,c}}{2\pi},\quad S_{\rm b,\,c}=\frac{\mathcal{A}_{\rm b,\,c}}{4G_N}~,
\end{equation}
where $\kappa_{\rm b,\,c}$ and $\mathcal{A}_{\rm b,\,c}=\Omega_{d-1}r^{d-1}_{\rm b,\,c}$ denote the surface gravities and areas of $\rm b$ and $\rm c$ respectively. For the surface gravity, we employ the same normalizations as in \cite{Bousso:1996au} which guarantee that the black hole and cosmological horizons are in thermal equilibrium for the Nariai configuration, {where} the inverse temperature (\ref{eq:thermo SdS}) in the Nariai limit is $\beta_{\rm N}=\frac{2\pi\ell}{\sqrt{d}}$.

\subsection{Preparation of the black hole}\label{sec:preparation BH}
We now discuss the conditions to produce the near-Nariai black hole by collecting radiation in a thermal bath in dS space. We model the bath as a Dyson sphere (i.e. a boundary wall that absorbs radiation) available to each of the static patch observers collecting the radiation coming from the cosmological horizon before sending the message. When trapping all the radiation into a box (i.e. the Dyson sphere) of size $r_{\rm box}\leq r_{\rm N}$, one can compute the time it would take to build a reservoir of any energy $E_{\rm res}\leq M_{\rm N}$ using Stefan-Boltzmann law in $(3+1)$-dimensions, 
\begin{equation}
    \mathcal{P}_{\rm c}=\sigma \mathcal{A}_{\rm c}\,T_{\rm c}^4
\end{equation}
with $\mathcal{P}_{\rm c}=\rmd M/\rmd t$ the radiated power, $\sigma$ the Stefan-Boltzmann constant, and $\mathcal{A}_{\rm c}$ the surface area of the cosmological horizon. Inserting the cosmological temperature, as well as the mass written in terms of the horizon radius $r_{\rm c}$
\begin{equation}
    T_{\rm c}=\frac{r_{\rm c}^2/r_{\rm N}^2-1}{4\pi r_c \sqrt{1-(M/M_{\rm N})^{2/3}}}\,,\quad M=\frac{r_{\rm c}}{2G_N}\left(1-\left(\frac{r_{\rm c}}{\ell}\right)^2\right)~,
\end{equation}
allows {us} to find
\begin{align}
        \frac{\rmd r_{\rm c}}{\rmd t}&=\frac{G_N \sigma  \left(\ell^2-3 r_{\rm c}^2\right)^3}{8 \pi ^3 \left(2 \ell^2 r_{\rm c}-3 \sqrt[3]{2} r_{\rm c}^{5/3} \left(\ell^2-r_{\rm c}^2\right){}^{2/3}\right)^2}\equiv \dot{r}_{\rm c}\,.
\end{align}
The time, $\tilde{\mathcal{T}}_{\rm N}$, necessary to fill up the whole dS space (till its radius matches the Nariai one) is therefore given by 
\begin{equation}
\tilde{\mathcal{T}}_{\rm N}=\int^{r_{\rm N}}_{\ell} \frac{\rmd r_{\rm c}}{\dot{r}_{\rm c}}\approx \frac{30\sqrt{\pi}}{G_N}r_{\rm N}^3\,,
\end{equation}
where we inserted $\sigma=\frac{2\pi^5}{15}$. The evolution of the cosmological and possible\footnote{In order to form a black hole, the energy density must be sufficiently large. It would require careful tuning of the position of the reflective boundary. However, we avoid any of these difficulties by filling up a box of size $r_{\rm N}$, guaranteeing that we have a black hole by the end of the process.} black hole radii in this process are illustrated in Fig. \ref{fig:dysonplot} (a).

\subsection{Shockwave preparation}\label{sec:preparation shockwaves}
\begin{figure}[t!]
    \centering
    \includegraphics[width=0.65\textwidth]{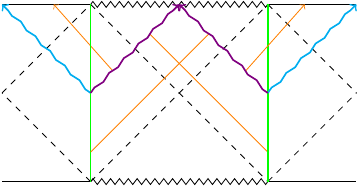}
    \caption{Configuration where the antipodal observers (in green) produce a thermal bath from Hawking radiation that they initially stored in the empty dS preparation of state. Each observer releases a positive energy pulse (cyan) to the inflating region, which also results in a negative energy pulse to the black hole interior (purple). This allows the observers to transmit light signals (orange) to each other.}
    \label{fig:bath}
\end{figure}

To produce a traversable wormhole in a near-Nariai black hole spacetime, the distance between the observer and the horizon has to \emph{grow}. For the inflating patch, this implies that the cosmological horizon radius has to increase, which is only achieved by an excess of positive energy leaking out of the considered patch or equivalently increasing its negative energy density. This implies that we send negative energy in the opposite direction, into the black hole's interior. See Fig. \ref{fig:bath} for an illustration of this step in the protocol.\footnote{Throughout the text, we employ the convention for the Penrose diagrams with shockwave shifts in which light signals are discontinuous along a path with the (cosmological or black hole) horizons remaining continuous.}
The deficit of positive or negative energy particles from one patch implies that there would be an excess of opposite sign energy in the opposite direction.

\begin{figure}[t!]
\centering
\begin{subfigure}[t]{0.485\textwidth}
    \includegraphics[width=\textwidth]{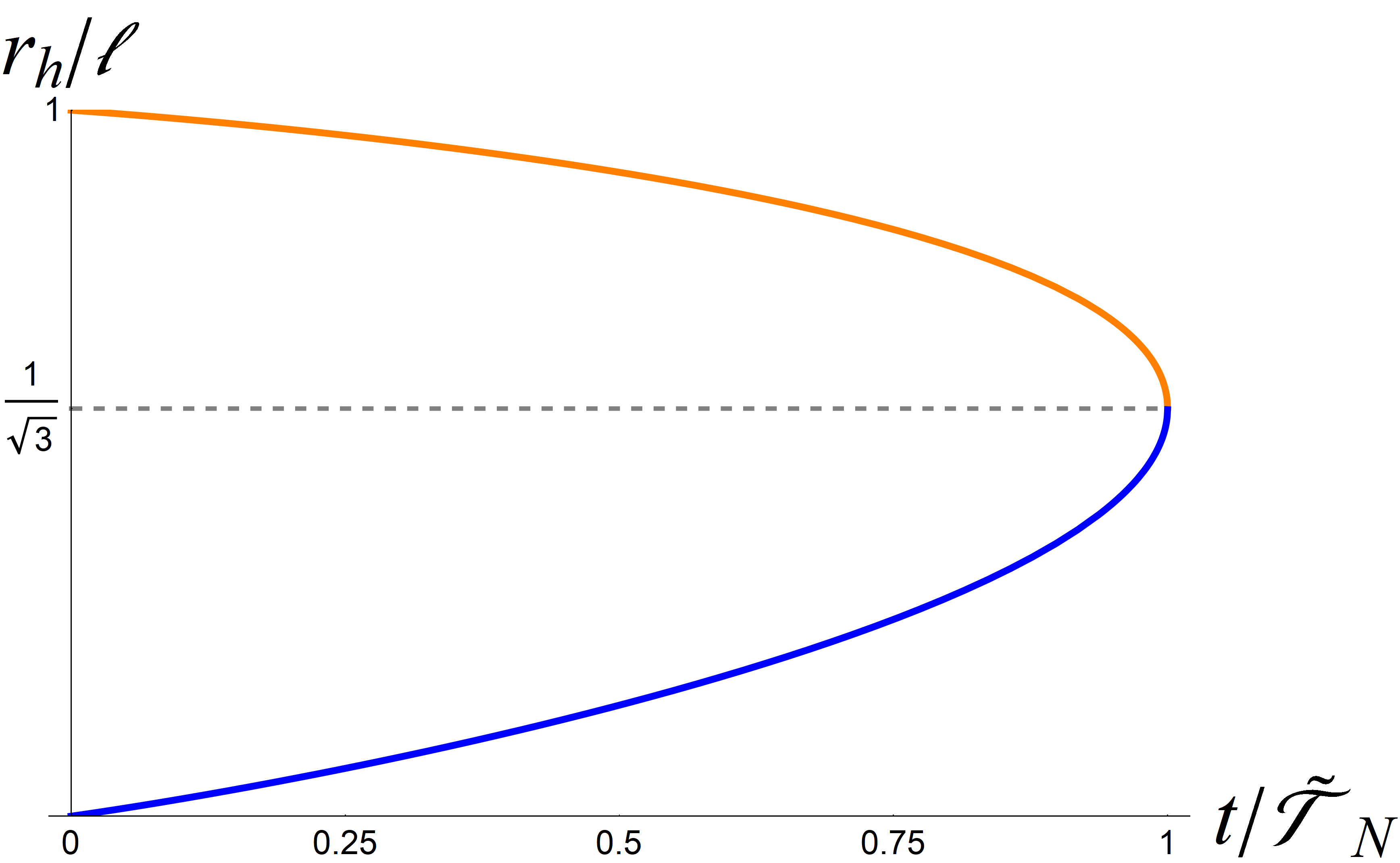}
    \caption{ }
\end{subfigure}\hspace{0.1cm}\begin{subfigure}[t]{0.485\textwidth}
    \includegraphics[width=\textwidth]{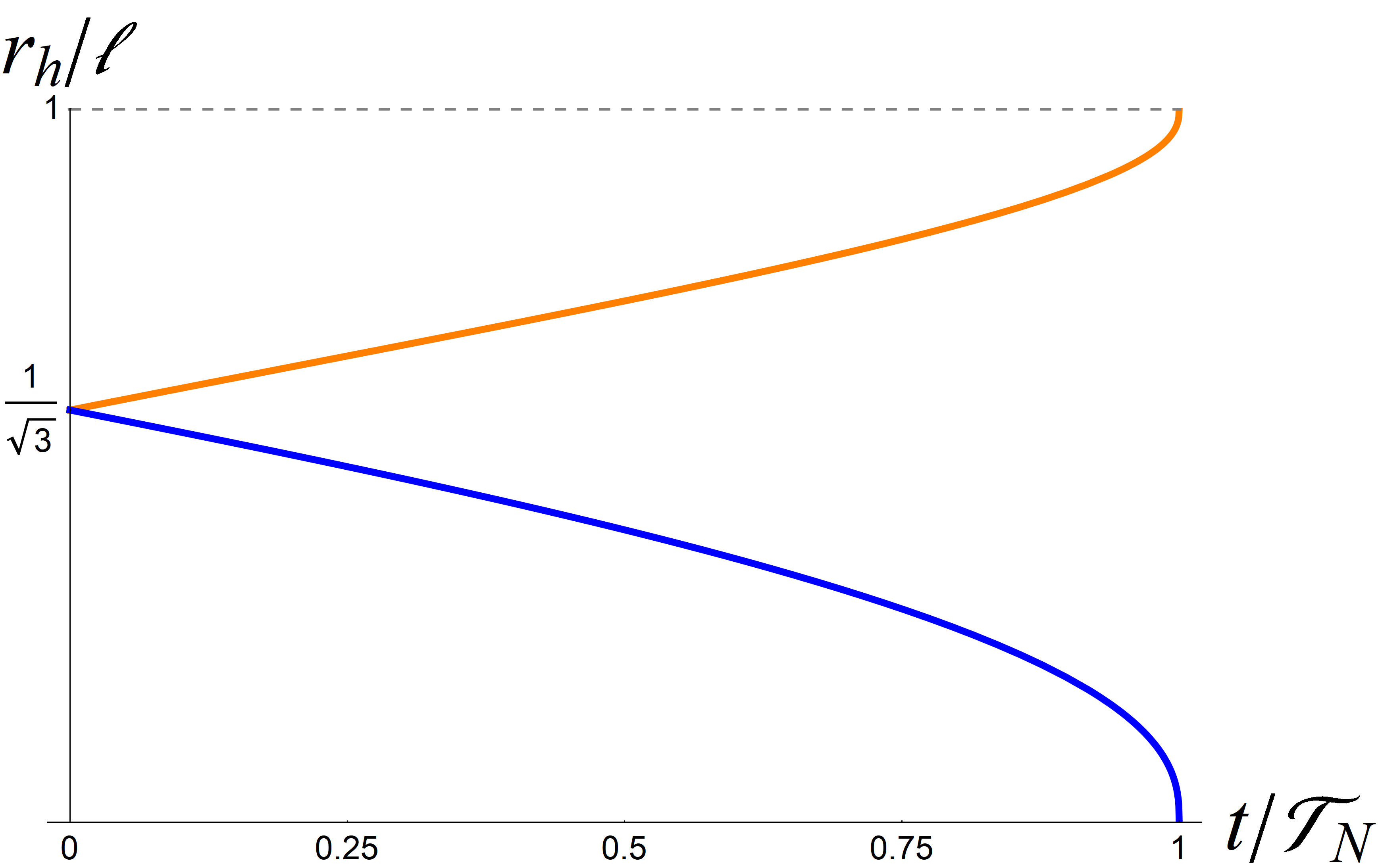}
    \caption{ }
\end{subfigure}
\caption{The horizon radii $r_h$ as a function of the static patch time in $d=3$. The cosmological radius $r_{\rm c}$ is shown in orange, and the black hole horizon radius is in blue. (a) We start from empty dS space and consider absorbing boundary conditions for trapping the radiation.  {T}he relation $\ell^2=r_{\rm b}^2+r_{\rm c}^2+r_{\rm b}r_{\rm c}$ allows us to plot the black hole radius that could be formed with the gathered energy. (b) We start with a Nariai black hole configuration and let the radiation freely escape through the cosmological horizon.}
\label{fig:dysonplot}
\end{figure}

We can treat this step of the protocol as in the previous section, but now the black hole and cosmological fluxes are not restricted by any reflective boundary or Dyson sphere. For a general SdS black hole, the total power passing through the cosmological horizon equals
\begin{equation}
\mathcal{P}_{\rm tot}=\mathcal{P}_{\rm c}-\mathcal{P}_{\rm b}\,.
\end{equation}
 The lifetime of a four-dimensional Nariai black hole, $\mathcal{T}_{\rm N}$, can be computed in a similar way as Sec. \ref{sec:preparation BH}, resulting in
\begin{equation}
    \mathcal{T}_{\rm N}\approx \frac{\tilde{\mathcal{T}}_{\rm N}}{\sqrt{2}}\,.
\end{equation}
The respective evolution is shown in Fig. \ref{fig:dysonplot} (b).

In the strict Nariai limit $\mathcal{P}_{\rm tot}=0$. However, when the black hole mass is very close to the Nariai mass $M=M_{\rm N}(1-\epsilon_{\rm N})$, where $0<\epsilon_{\rm N}\ll1$, one can find that at first order
\begin{align}
    \mathcal{P}_{\rm tot}(M\rightarrow M_{\rm N})&=\mathcal{P}_{\rm tot}(M_{\rm N})+ \epsilon_{\rm N} \left(\frac{\partial \mathcal{P}_{\rm tot}}{\partial \epsilon_{\rm N}}\right)_{\epsilon_{\rm N}=0}=\frac{\epsilon_{\rm N}}{3\pi^3 \ell^2}\,.
\end{align}
Then, the energy $E_{\rm N}$ released when opening the box for a duration $\delta\tau$ for a SdS black hole close to the Nariai limit is given by
\begin{equation}
    E_{\rm N}=\frac{\epsilon_{\rm N}}{3\pi^3\ell}\left(\frac{\delta\tau}{\ell}\right)\,.
\end{equation}
Such emission of energy can cause a run-away effect leading to the complete evaporation of the black hole. Notice that imposing $\delta\tau/\ell\equiv\epsilon_{\tau}\ll 1$, the released energy ratio $E/M$ by the pulse in the Nariai limit can be approximated by
\begin{equation}\label{eq:near-Nariai energy}
    \frac{E_{\rm N}}{M_{\rm N}}\approx \epsilon_{\rm N}\,\epsilon_{\tau} \left(\frac{\ell_{\rm pl}}{\ell}\right)^2\,.
\end{equation}
The result is an extremely small quantity, therefore justifying a shockwave description.

\subsection{GJW protocol for Nariai spacetimes}
As the observers throw matter, they are able to generate a traversable wormhole in the black hole interior, as well as in the inflating region. In that way, the observers can transmit light signals through the wormhole, as shown in Fig. \ref{fig:bath}.\footnote{Operators supported in previously causally disconnected spacetime points in the static patches of the black hole and cosmological regions become causally connected. This modifies the commutation relations of the algebra of observables \cite{Chandrasekaran:2022eqq}. We thank Aron Wall for important comments about this point.} Notice that we do not require a holographic theory to establish this step in the protocol.\footnote{A key difference with respect to realizing the protocol in an asymptotically AdS black hole spacetime is that we do not need to violate ANEC in order to open the wormhole, as we have elaborated. We thank Ben Freivogel for his useful question about this point.}

To make the previous statements more precise, we calculate below the bounds on the amount of information transmitted during the process.
\begin{itemize}
    \item For the \textbf{black hole patch}, we focus on the static patch time that the message spends behind the horizon of Fig \ref{fig:bath}, given by
\begin{equation}
    \Delta t = t_{\rm lc}+t_{\rm shift}+\delta t \, ,
\end{equation}
where $t_{\rm lc}$ stands for the light-crossing time, $t_{\rm shift}$ corresponds to the time for gathering the required energy and $\delta t$ is the time delay between collecting the energy and sending the message.

For the gathering time, $t_{\rm shift}$, we consider the reflection time for a photon to go from the radius of the Nariai black hole ($r=r_{\rm N}$) to the location of the static patch observer at $r=0$, which is given by:
\begin{equation}
    t_{\rm shift}\approx \ell \, \log{S_{\rm b}}~,
\end{equation}
where $S_{\rm b}=S_{\rm N}/2$ is the Nariai black hole entropy. In the following, we consider the minimum possible time for information recovery, where $t_{\rm lc}$ and $\delta t$ can be neglected. This implies that the time difference before and after the absorption of photons, where the black hole entropy increases due to the presence of the shockwave as $S_{\rm b}\xrightarrow{}S_{\rm b}+S_{\rm Shock}$, is given by 
\begin{equation}
    \Delta t\propto\log[S_{\rm b}+S_{\rm Shock}]~.
\end{equation}
In that case, the recovery time is given as
\begin{align}\label{eq:Time scales BH}
   \Delta\tau&\equiv\Delta t-\tau^{*}_{\rm b}\propto\log\qty[1+\frac{S_{\rm Shock}}{S_{\rm b}}]~,
\end{align}
where $\tau^{*}_{\rm b}= \ell\log S_{\rm b}$ is the black hole scrambling time.

The condition for the wormhole to be traversable requires the recovery time to be smaller than the original scrambling time ( $\Delta\tau<0$), such that
\begin{equation}
    -S_{\rm b}<S_{\rm Shock}<0~.
\end{equation}
The result agrees with the fact that the shock wave should have the opposite sign than the energy of the black hole, as this kind of shock-wave \emph{reduces} its entropy.

\item For the \textbf{inflating patch}, we may use the same expression as before, $\Delta t = t_{\rm lc}+t_{\rm shift}+\delta t$, such that, neglecting the contribution of $\delta t$, the recovery time is now given as follows
\begin{equation}
\begin{aligned}
       \Delta t&= t_{\rm lc}+t_{\rm shift}\propto\log[S_{\rm c}-S_{\rm Shock}]~ .
\end{aligned}
\end{equation}
Therefore, the difference in emission times is then
\begin{equation}\label{eq:time scale CH}
\begin{aligned}
    \Delta\tau&=\Delta t-\tau^{*}_{c}\propto\log\qty[1-\frac{S_{\rm Shock}}{S_{\rm c}}]~.
\end{aligned}
\end{equation}
with  $\tau^{*}_{c}= \ell\log S_{\rm c}$. We again notice that $\Delta\tau<0 $ is necessary for a traversable wormhole, which requires
\begin{equation}
    0<S_{\rm Shock}<S_{\rm c}~.
\end{equation}
This result reflects the fact that the shockwave has to increase the entropy of the cosmological horizon to generate the traversable wormhole. It also corroborates that the Gibbons-Hawking entropy is counting the degrees of freedom behind the horizon.
\end{itemize}

 As argued in \cite{Freivogel:2019lej, Freivogel:2019whb}, regardless of the protocol, the maximal amount of information that can be transmitted through the wormhole in this context is bounded by
\begin{equation}\label{eq:N bits}
    N_{\rm bits}\lesssim r_{\rm N}^{d-1}/G_N\approx S_{\rm b}~,
\end{equation}
given that we cannot transfer more bits of information than those available from the surface area of the black hole horizon, as it gets destroyed during the process. See App. \ref{App:HaydenPreskill} for an explanation of the relation between the number of bits and the criterion for successful information recovery. 
 This suggests that the amount of information passing through the resulting wormhole throats in SdS space can be estimated by considering the differences
\begin{align}
    \Delta S_{\rm b}&=S_{\rm b}(M)-S_{\rm b}(M-E)\\
    \Delta S_{\rm c}&=S_{\rm c}(M-E)-S_{\rm c}(M)\,,
\end{align}
where $E=\int \mathcal{P}_{\rm tot}~\rmd t$ is the total energy exchanged during the protocol. Notice that the maximal total information that can be transmitted in this way is given by
\begin{equation}\label{eq:Main bound}
    \Delta S_{\rm tot}=\Delta S_{\rm b}+\Delta S_{\rm c}=S_{\rm N}/2+(S_{
\rm dS}-S_{\rm N}/2)=S_{\rm dS}~,
\end{equation}
with $E=M_{\rm N}$; whereas the total entropy difference {is} $\Delta S=S_{\rm dS}-S_{\rm N}=S_{\rm N}/2$ in (3+1)-dimensions. Furthermore, assuming small energy ratios $E/M\equiv\epsilon$ and expanding at first order, we find that
\begin{align}
    s_{\rm b}&\equiv\epsilon \left(\frac{\partial S_{\rm b}}{\partial \epsilon}\right)_{\epsilon=0}=\epsilon M \frac{\partial S_{\rm b}}{\partial M}= E \beta_{\rm b}~,\\
    s_{\rm c}&\equiv-\epsilon \left(\frac{\partial S_{\rm c}}{\partial \epsilon}\right)_{\epsilon=0}=-\epsilon M \frac{\partial S_{\rm c}}{\partial M}= E \beta_{\rm c}\,,
\end{align}
where we normalized the {inverse temperature} $\beta$ with respect to the observers at $r_{\mathcal{O}}$. Notice that the wormhole sizes do not coincide, i.e. $s_{\rm b}\neq s_{\rm c}$, for any $M$ except the Nariai mass $M_{\rm N}$, where $\beta_{\rm b}=\beta_{\rm c}=2\pi r_{\rm N}$. 
\begin{figure}[t!]
\centering
\begin{subfigure}[t]{0.485\textwidth}
    \includegraphics[width=\textwidth]{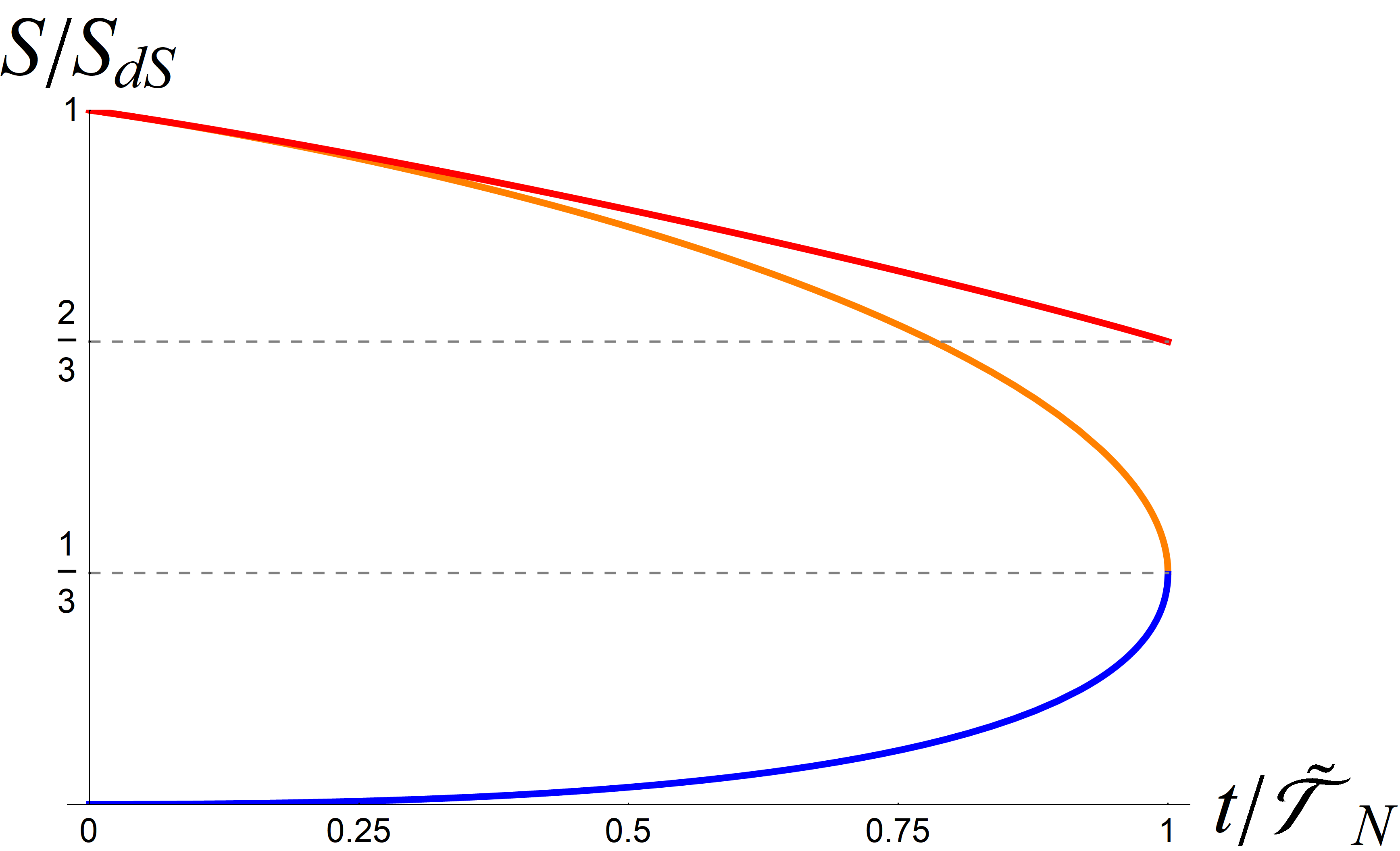}
    \caption{ }
\end{subfigure}\hspace{0.1cm}\begin{subfigure}[t]{0.485\textwidth}
    \includegraphics[width=\textwidth]{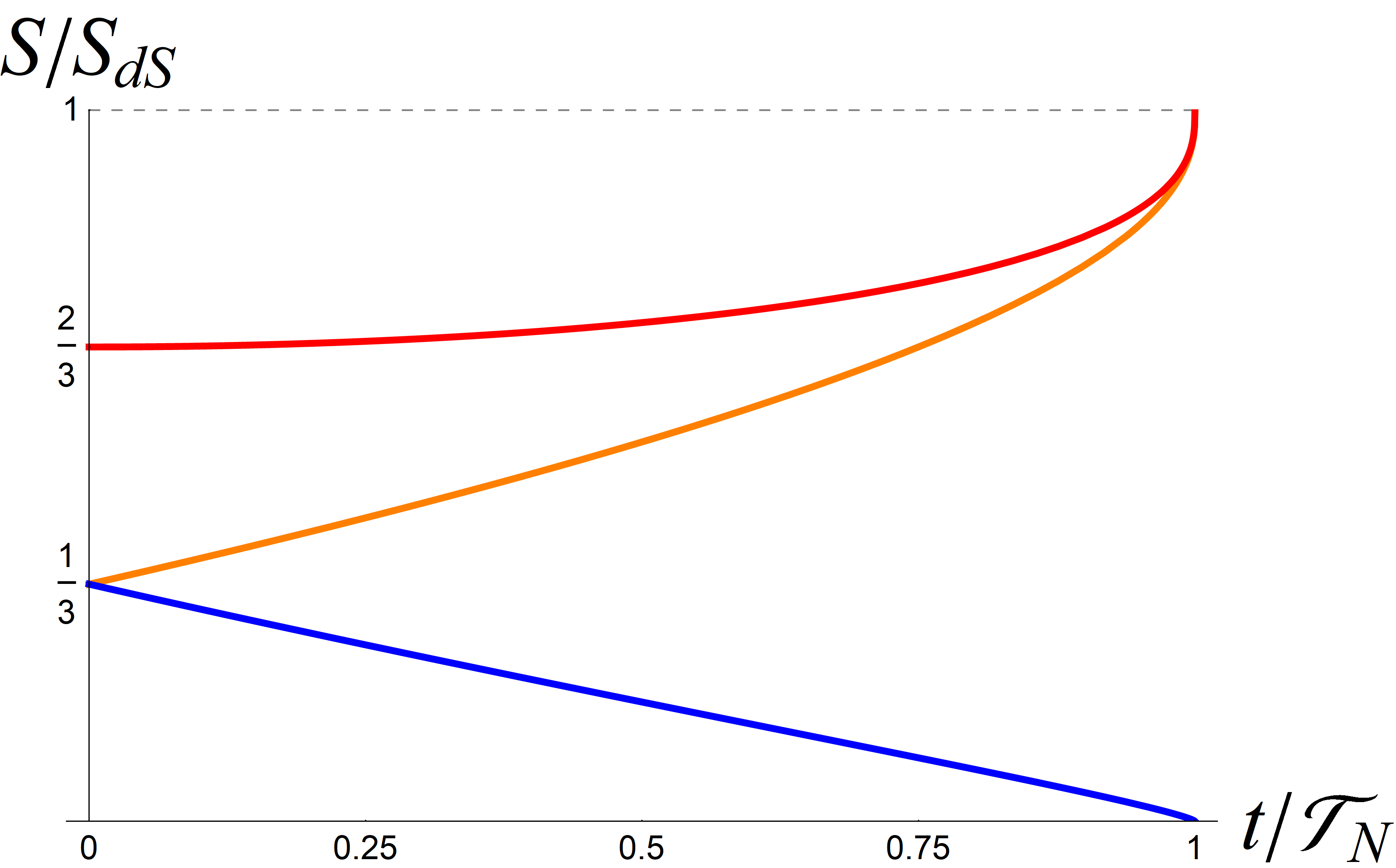}
    \caption{ }
\end{subfigure}
\caption{The cosmological entropy $S_{\rm c}$ (orange), black hole entropy $S_{\rm b}$ (blue), and their sum, the total entropy (red), as a function of time in $d=3$.
(a) Empty dS with absorbing boundary conditions from the Dyson sphere to capture the energy. (b) Nariai spacetime where the energy flux freely escapes through the cosmological horizon.}
\label{fig:dysonplotentro}
\end{figure}
Figure \ref{fig:dysonplotentro} shows the evolution of the entropies as a function of time.

\section{Information recovery in (1+1)-dimensions}\label{Sec:2d}
In this section, we incorporate semiclassical effects into our protocol. For that, we adopt dS JT gravity from the S-wave reduction of the near-Nariai black holes\footnote{The island rule has been derived in \cite{Svesko:2022txo} for the type of model under consideration, albeit for static backgrounds. However, our findings point out that the island still obeys maximin conditions \cite{Akers:2019lzs}, and the shockwave simply shifts its location.}. The {theory} is defined by the action
\begin{equation}\label{eq:JT action}
\begin{aligned}
    I_{\rm dS\,JT}=&\frac{\phi_0}{16\pi G_2}\int_{\mathscr{M}}\rmd^2x\sqrt{-{g}}R+\frac{1}{16\pi G_2}\int_{\mathscr{M}}\rmd^2x\sqrt{-{g}}\qty({R}-\frac{2}{\ell^2})\phi\\
    &+\frac{1}{8\pi G_2}\int_{\partial\mathscr{M}}\rmd y\sqrt{-{h}}\,\phi {K}+I_{\rm matter}~,
\end{aligned}
\end{equation}
where $\phi$ is the dilaton field, {which describes} scalar mode perturbations around the Nariai black hole solution, $G_2$ the effective gravitational coupling, {and $I_{\rm matter}$ denotes the bulk CFT$_2$ theory}. See appendix \ref{App:JT from SdS} for more details on deriving this theory from the dimensional reduction of a near-Nariai black hole. The equations of motion, {{corresponding to} (\ref{eq:JT action}), are given by}
\beq
\bal
 - \nabla_\mu\nabla_\nu\phi + g_{\mu\nu}\square\phi + \frac{\phi}{\ell^2}g_{\mu\nu}&= 8\pi G_2 \langle
 T_{\mu\nu}
 \rangle ~,\\
R&=2/\ell^2 ~,\label{eq:EOMJTdS}
\eal
\eeq
with $\expval{T_{\mu\nu}}$ the expectation value of the stress tensor.

\subsubsection*{Islands before the shockwave}
\begin{figure}[t!]
    \centering
        \includegraphics[width=0.65\textwidth]{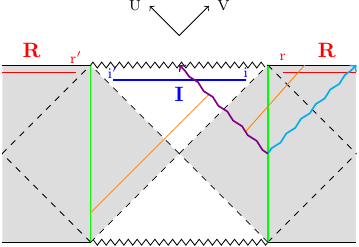}
\caption{Proposal for the teleportation protocol in the dS JT gravity. Same color labeling as in Fig. \ref{fig:bath}. The signals sent through the wormhole arrive at a pair of detectors, $\mathbf{R}$, in an interval bounded by the points $[r,\,r']$ near $\mathcal{I}^+$ of the inflating patch. The radiation is purified by the island, $\mathbf{I}$ (blue), {which is} bounded by the points $[i,\,i']$. The {axes} for the Kruskal coordinates ($U,\,V$) {are} also shown. The gray area represents the region covered by a single Kruskal coordinate system, which includes the static patches and either the black hole or the inflating regions.}
    \label{fig:Setup_Islands_detectors}
\end{figure}

We now place detectors near {the } $\mathcal{I}^+$ {region} of the inflationary patches. The metaobservers can then collect Hawking modes in weakly gravitating regions. See Fig. \ref{fig:Setup_Islands_detectors} for an illustration of the proposal. 
In the configuration of interest, we evaluate {the generalized entropy (\ref{eq:Sgen}) considering that the state on {a complete} Cauchy slice is pure at a given time},\footnote{Notice that in our protocol, only the global state of the bulk geometry together with the energy pulses is pure. This occurs since we prepared the Bunch-Davis state in pure dS space, and later trapped some of the matter fields into a reservoir carried by the antipodal observers.} {the state on $\mathbf{R}\cup\mathbf{I}$ is purified by that on the complement region, resulting in} $S_{\rm gen}(\mathbf{R}\cup\mathbf{I})=S_{\rm gen}(\overline{\mathbf{R}\cup\mathbf{I}})$. We consider a completely symmetric configuration for the endpoints of the island and those of the detectors in $\mathbf{R}$. {Let us introduce }conformal vacuum coordinates $x^\pm$. {The metric has the form }
\begin{equation}
\rmd s^2 =- \Omega(x^+,\,x^-)^{-2}\rmd x^+\rmd x^- ~,
\end{equation}
where we denote $\Omega(x^+,\,x^-)$ as the conformal factor.

The von Neumann entropy for a single interval $[j,\,k]$ is given by
\beq\label{eq:VNE}
S_{\rm vN}([j,\,k])= \frac c6 \log\left[\frac{(x^+_j-x^+_k)(x^-_j-x^-_k)}{\Omega(x^+_j,\,x^-_j)\Omega(x^+_k,\,x^-_k)\epsilon_j\epsilon_k}\right] ~,
\eeq
where the factors $\epsilon_j$ and $\epsilon_k$ denote the usual short-distance cutoffs of the von Neumann entropy. Meanwhile, the dilaton $\phi(\partial I)$ gives the area contribution in (\ref{eq:Sgen}),
\begin{equation}
   S_{\rm gen}(\mathbf{R}\cup\mathbf{I})=\frac{\phi(\partial \mathbf{I})}{4G_N}+S_{\rm matter}(\mathbf{R}\cup\mathbf{I})~.
\end{equation}
We use Kruskal coordinates ($V,\,U$) to denote the metric and the \emph{sourceless} dilaton
\begin{equation}\label{eq:Metric Kruskal}
\begin{aligned}
        \rmd s^2&=-\frac{4\ell^4}{(\ell^2-UV)^2}\rmd U\,\rmd V~,\\
    \phi(U,\,V)&=\phi_r\frac{\ell^2+UV}{\ell^2-UV}~.
\end{aligned}
\end{equation}
Let $(U_i,\,V_i)$ denote one of the endpoints of $\mathbf{I}$, and $(U_r,\,V_r)$ one of the endpoint in $\mathbf{R}$, as shown in Fig. \ref{fig:Setup_Islands_detectors}. Islands in the Bunch-Davis vacuum dS JT gravity have been derived in many {places throughout} of the literature in the limit that the spatial interval distance of regions $\mathbf{R}$ and $\mathbf{I}$ is large compared to the distance between separating them (see \cite{Hartman:2020khs, Svesko:2022txo} for example). We need an analytic continuation in the Kruskal coordinate system to evaluate the von Neumann entropy given that $\mathbf{I}$ and $\mathbf{R}$ are located in the black hole and inflating patches. {T}he metric (\ref{eq:Metric Kruskal}) in global conformal coordinates {is given by}
\begin{equation}\label{eq:global coord}
    \rmd s^2=\ell^2\sec^2\sigma~(-\rmd\sigma^2+\rmd\varphi^2)~.
\end{equation}
{Now, o}ne needs to make a shift of coordinates $\varphi\xrightarrow{}\varphi+\pi$ to connect points in the adjacent hyperbolic patches. This amounts to shifting the point $(V_r,\,U_r)$ as
\begin{equation}\label{eq:cont hyperbolic}
    V_r\xrightarrow{}-\frac{\ell^2}{V_r},\quad U_r\xrightarrow{}-\frac{\ell^2}{U_r}~.
\end{equation}
For the region below the shockwave in Fig. \ref{fig:Setup_Islands_detectors}, the generalized entropy can be then found as
\begin{equation}
S_{\rm gen}=\frac{1}{2G_2}\qty(\phi_0-\phi_r\frac{\ell^2+U_iV_i}{\ell^2-U_iV_i}+\frac{cG_N}{3})+\frac{c}{3}\log\frac{4\ell^2(\ell^2+U_iU_r)(\ell^2+V_iV_r)}{\epsilon_{i}\epsilon_{r}(\ell^2-U_iV_i)(\ell^2-U_rV_r)}~.
\end{equation}
The only island space-like separated from $(V_r,\,U_r)$ is given by
\begin{equation}\label{eq:island vanilla}
\begin{aligned}
    V_i&=\frac{U_r\epsilon}{3}+\mathcal{O}(\epsilon^2)~,\quad U_i=\frac{V_r\epsilon}{3}+\mathcal{O}(\epsilon^2)~,
\end{aligned}
\end{equation}
where $\epsilon\equiv\frac{G_2c}{\phi_r}$ is a small parameter in the semiclassical regime, {i.e.} $\epsilon\ll1$.

However, in the extremization of the generalized entropy (\ref{eq:Sgen}) one must also include the possibility for the island to be the empty set, $\emptyset$. In that case, we can produce a Page curve transition between the trivial and non-trivial island (\ref{eq:island vanilla}) saddles. One can derive that this transition occurs in global coordinates (\ref{eq:global coord}) for a detector in the interval $[r, r']$ for $\varphi_r=-\varphi_{r'}\equiv \varphi_{\rm Page}$ and $\sigma_R\simeq\frac{\pi}{2}$ constant in Fig. \ref{fig:Setup_Islands_detectors}, {and it} is given by \cite{Hartman:2020khs}
\begin{equation}\label{eq:phi Page}
    \tan\varphi_{\rm Page}\propto \exp[\frac{3}{2G_2c}\qty(\phi_0+\phi_r+\frac{cG_2}{3})]~.
\end{equation}

\subsubsection*{Adding the shockwaves}
{Now,} we allow the observers to release negative energy pulses toward the black hole's interior. Consider for instance the south pole observer in Fig. \ref{fig:Setup_Islands_detectors}. The stress tensor is given by\footnote{It is possible to set a more physical situation in which the stress tensor becomes a smooth function. This can be done by adding a smearing function $f(V)$, {and consider} 
\be
\langle \tilde{T}_{VV} \rangle = \int dV \langle T_{VV} \rangle f(V)~. 
\ee
In general, we expect that the specific form of smearing function depends on the details of the theory. The resulting effect will cause the stress tensor not to change the geometry abruptly \cite{Freivogel:2018gxj}. } 
\beq\label{eq:stress tensor non equil}
\bal
\langle T_{VV}(V)\rangle &= \alpha\frac{\delta(V-V_S)}{V_S}~,
\eal
\eeq
which describes a shockwave moving towards the black hole interior along the null ray $V=V_S$. As in Sec. \ref{sec:Set up}, we employ $\alpha<0$ to form a traversable wormhole on the black hole interior, and we consider that $\abs{\alpha}\ll1$. See further comments on the small shift approximation in App. \ref{App:Shift SdS}.

{Next,} we would like to compare the island location with the shockwave insertion in the protocol, to inquire under what conditions the traversable wormhole enters in causal contact with the island region. Given the stress tensor (\ref{eq:stress tensor non equil}), the dilaton solution in (\ref{eq:Metric Kruskal}) would be modified according to the graviton equation of motion {in} (\ref{eq:JT action}), {leading to}
\begin{equation}\label{eq:backreacted dilaton}
    \phi=\phi_0-\phi_r\frac{\ell^2+U_iV_i}{\ell^2-U_iV_i}+\frac{cG_2}{3}-\frac{8\pi G_2\alpha}{V_S}\frac{\ell^2-V_SU_i}{\ell^2-V_iU_i}(V_i-V_S)\theta(V_i-V_S) ~.
\end{equation}
We want to identify null congruences, which in JT gravity are described by curves satisfying $\partial_{\pm}\Phi=0$ \cite{Aalsma:2021kle}. We denote them as follows
\begin{align}
\gamma:\quad V&=V_S\frac{2G_2\alpha}{\phi_r}+\mathcal{O}(\alpha^3)~,\\
    \delta^+:\quad U&=\frac{\ell^2}{V_S}\frac{\phi _r}{2G_2\alpha}+\mathcal{O}(\alpha),\\
    \delta^-:\quad U&=-\frac{\ell^2}{V_S}\frac{2G_2\alpha}{\phi_r}+\mathcal{O}(\alpha^3)~.\label{eq:Delta U}
\end{align}
The apparent horizon is the outermost surface, with $\partial_+\Phi = 0$ and $\partial_-\Phi < 0$ \cite{Aalsma:2021bit}, which is allowed by both of the curves $\delta^\pm$. The curve describing the null shift for $\alpha<0$, which we implement in the protocol to violate ANEC, corresponds to the one that reduces the size of the black hole, which is $\delta^-$ for $V>V_S$. This means that the null shift of the apparent horizon to leading order in $\alpha$ is given by (\ref{eq:Delta U}).

In our conventions, the horizon in the Penrose diagrams remains fixed while the light signals are shifted. As the observer in the north pole sends a light signal to the causal diamond of the island $\mathbf{I}$ in Fig. \ref{fig:Setup_Islands_detectors}, it will get shifted by (\ref{eq:Delta U}), and it will {reach} the location $\mathbf{R}$ of the adjacent hyperbolic slice. The signal sent along $U=U_{\rm ls}$ enters the inflating patch along the line:
\begin{equation}\label{eq:null shift}
    U=U_\alpha\equiv\frac{\ell^2}{U_{\rm ls}}+\frac{8\pi G_2\abs{\alpha}}{4\pi\phi_r}~,
\end{equation}
{which follows from} the analytic continuation {in} (\ref{eq:cont hyperbolic}). Then, the light ray would intercept the causal diamond of the island, found in (\ref{eq:island vanilla}), as long as $U_r<U_\alpha$.

\subsection*{Repeating the protocol}
As we noticed in (\ref{eq:near-Nariai energy}), the energy released as shockwaves in our protocol is generically small, as it is suppressed by a $\ell_{\rm pl}^2$ factor. The static patch observers can then decide on performing multiple shockwave protocols, as long as it obeys the bound found in (\ref{eq:Main bound}). In this situation, one must account for backreaction on the island location. Given the presence of the stress tensor in (\ref{eq:stress tensor non equil}), one must find the map to vacuum coordinates $x^\pm$ in (\ref{eq:VNE}). We can find such a map using the conformal anomaly of the stress tensor
\begin{equation}
    \expval{T_{VV}(V)}=-\frac{c}{24\pi}\qty{x^+,\,V}~.
\end{equation}
Taking the Ansatz
\begin{equation}
    x^+=V+\alpha\,f(V),\quad x^-=U~,
\end{equation}
for $\alpha\ll1$ we obtain the following relation at leading order in $\alpha$
\begin{equation}\label{eq:map f(xpm)}
    -\frac{c}{24\pi}f'''(V)=\frac{1}{V_S}\delta(V-V_S)~,
\end{equation}
{where} the solution {to the above differential equation} is found to be
\begin{equation}
    f(V)=-\frac{12\pi}{c~V_S}(V-V_S)^2\Theta(V-V_S)~.
\end{equation}
Consider the vacuum coordinate above, for the von Neumann entropy in (\ref{eq:VNE}) and the backreacted dilaton in (\ref{eq:backreacted dilaton}). We will be interested in the regime
\begin{equation}\label{eq:Regime eps alp}
    \epsilon\ll\abs{\alpha}\ll1~.
\end{equation}
The extremization of the generalized entropy (\ref{eq:Sgen}) leads to the shifted island location
\begin{align}
    V_i&=\frac{U_r\epsilon}{3}+16\pi G_2\alpha\frac{V_S\epsilon}{c}+\mathcal{O}(\epsilon^2,~\alpha^2\epsilon)~,\label{eq:Ansatz alpha epsilon}\\
    U_i&=\frac{V_r\epsilon}{3}-16\pi G_2\alpha\frac{\epsilon}{cV_S}+\mathcal{O}(\epsilon^2,~\alpha^2\epsilon)~,\label{eq:xI-}
\end{align}
for which, the von Neumann entropy collected on $\mathbf{R}$ due to the island is given as
\begin{equation}\label{eq:Sisland}
\begin{aligned}
    S_{\rm isl}=&\frac{\phi_0}{G_2}+\frac{c}{6}\qty(1-\frac{3}{\epsilon}+2\log\frac{4\ell^4}{\epsilon_{i}\epsilon_{r}(\ell^2-U_rV_r)})\\
    &+16\pi G_2\alpha+\frac{\epsilon\qty(c V_SV_rU_r+48\pi G_2\alpha\qty(V_S^2V_r-\ell^2U_r))}{9\ell^2V_S}~.
    \end{aligned}
\end{equation}
The result agrees with \cite{Hartman:2020khs,Svesko:2022txo} when $\alpha=0$.

We now consider the situation illustrated in Fig. \ref{fig:Setup_Islands_detectors} where $V_rU_r\approx\ell^2$ (corresponding to $\mathbf{R}$ being close to $\mathcal{I}^\pm$). (\ref{eq:Sisland}) to $\mathcal{O}(\alpha)$ in the regime (\ref{eq:Regime eps alp}) becomes
\begin{equation}\label{eq:Sisl const}
    S_{\rm isl}\simeq S_{\rm BH}+\frac{c}{3}\log\frac{4}{\epsilon_i\epsilon_r\delta L^2}+16\pi G_2\alpha~,
\end{equation}
where $\delta L^2=\ell^2-U_rV_r$, and $S_{\rm BH}=\frac{1}{2G_2}\qty(\phi_0-\phi_r)$ is the Bekenstein-Hawking entropy of the Nariai black hole. Notice that (\ref{eq:Sisl const}) provides {a} bound on the amount of information that can be transmitted between the observers through the island. The classical regime corresponds to $S_{\rm isl}\simeq S_{\rm BH}$, similar to (\ref{eq:N bits}). However, notice that we are not comparing information bounds due to the opening of the traversable wormhole in the SdS$_4$ geometry in Sec. \ref{sec:Set up} with the von Neumann entropy derived from the island rule in the setup shown in Fig. \ref{fig:Setup_Islands_detectors}. {The former} relies on a classical bulk geometry in 4-dimensional gravity, while the latter {relies} on a semiclassical theory whose matter content is inherently 2-dimensional, {despite the fact that} the geometry comes from the s-wave reduction of a higher dimensional near-Nariai black hole spacetime.

As we repeat the procedure, the total null shift of the light rays intercepted by the north pole observer in Fig. \ref{fig:Setup_Islands_detectors} would then correspond to (\ref{eq:null shift}) with $\alpha\rightarrow\alpha'<0$ for the net null shift, and the condition of intercepting the island corresponds to $U_r<U_{\alpha'}$.

\section{Discussion and outlook}\label{Sec:Discussion}
In summary, we proposed a new protocol to form traversable wormholes in SdS black holes. Due to the lack of a holographic description, a conceptual difficulty in the protocol arises. Consider generating a traversable wormhole without adding a coupling between antipodal static patch observers in the near-Nariai black hole. It might seem that one can produce a signal between causally disconnected regions when there is no proper synchronization of the observers to emit the shockwaves responsible for the traversable wormhole. For this reason, we started from the Bunch-Davis state of dS space as the initial configuration. The antipodal observers are then maximally entangled, and by employing a classical communication channel, their clocks can then be synchronized. We also require that the pair of observers store energy coming from the Hawking radiation of the cosmological horizon, so they can later collapse the energy to form a near-Nariai black hole. {Later}, we proposed a symmetric protocol, where each of the observers throws an excess of positive energy into the cosmological horizon, which results in a negative energy excess into the black hole horizon. Then, a traversable wormhole will form in the black hole interior, as well as in the inflationary region. Each observer can send light signals at the appropriate time, {that can} cross through the newly formed wormhole. {Thus,} they are able to apply an appropriate recovery channel and decode the information by an appropriate protocol. In order to quantify such recovery, we analyzed bounds on the amount of information that can be transmitted through the wormhole. In the second part of the manuscript, we incorporated semiclassical effects explicitly by establishing the protocol in dS JT gravity and probing the island region. We considered how information can be transmitted through a wormhole in the island according to the metaobserver in the inflating region. Notice that our protocol does not advocate a particular notion of dS holography, and {it} might as well provide new hints towards a yet-to-be-known holographic dual theory to the SdS space-time. In addition, when not restricted to a shockwave description ($E/M\ll 1$), the maximal amount of bits that can be transferred from one static patch to the other is equal to the total number of bits contained in \textit{empty} dS space. Naively, one could have guessed that solely the entropy of the black hole would provide a bound on the possible information exchange ($\Delta S_b=S_N={\rm Area}(r_N)/(4G_N)$). However, releasing $M_N$ of energy will not only destroy the Nariai black hole but {it will also} extend the cosmological horizon as well (since $\Delta S_c=S_{dS}-{\rm Area}(r_N)/(4G_N)$).

We now comment on some extensions of our work.

\subsection*{Other extensions with the island formula}
One of the main drives in this work was incorporating bulk matter in the background to probe the island region with our protocol. This was explicitly performed in $(1+1)$-dimensions, but we expect it can be similarly done in higher dimensions. To study such effects in an explicit setting, perhaps double holographic models with dS braneworlds \cite{Aguilar-Gutierrez:2023tic,Aguilar-Gutierrez:2023zoi} would be a fruitful {future} direction. One might be able to incorporate quantum SdS black holes \cite{Emparan:2022ijy,Panella:2023lsi} in such a setting to treat quantum backreaction effects exactly.

Moreover, the 2-dimensional analysis was based on the s-wave reduction of a near-Nariai black hole configuration with shockwave sources, to simplify the evaluation. However, given the instability of this configuration, one should in principle describe the evaporation of arbitrary dS black holes from dS JT gravity. Modeling the evaporation of dS black holes in dS JT gravity has been made in \cite{Balasubramanian:2020xqf,Baek:2022ozg}. There has been a lot of progress in information recovery during an adiabatic evaporation regime with free particle bulk matter fields \cite{Hollowood:2021nlo} that could be extended to SdS space.

An alternative analysis of the islands could be {pursued} with the set-up of \cite{Aalsma:2022swk}. In {that} model, the radiation is collected in a subregion of asymptotically flat space coupled to the dS JT gravity model, instead of considering the detectors near $\mathcal{I}^+$ of the inflating patches, {as it was} studied here. It might be possible to reproduce our protocol in this setting with better analytic control over backreaction effects. In \cite{Aalsma:2022swk}, one can encounter the islands in the inflating patch of dS$_2$ space due to backreaction effects, which do not affect the observers in the asymptotically flat region. Perhaps this setting would allow us to study a traversable wormhole on the island in the inflating region which we have discussed in Sec. \ref{Sec:2d}.

\subsection*{States and algebra of observables}
In the more general case of a SdS black hole, there is a conceptual difficulty in properly defining states given that the cosmological horizon and the black hole horizon are not in thermal equilibrium. In this case, the construction of Hilbert spaces can enter into ambiguities for defining thermofield double states and studying the algebra of semiclassical states on the SdS background. It would be very interesting to make a rigorous treatment of quantum states of these types of black holes, based on recent developments in von Neumann algebras \cite{Jensen:2023yxy,Kudler-Flam:2023qfl,Faulkner:2024gst} (see also \cite{vanderHeijden:2024tdk} where an algebraic approach to a teleportation protocol in AdS black holes has been proposed). For our protocol, it might be possible to make a TFD-type state preparation of modes coming from the near-Nariai black hole and cosmological horizons, as well as the modes contained in the thermal baths. We leave a more rigorous treatment of the quantum states involved in our protocol for future directions. Moreover, it would be interesting to study whether the preparation of a vacuum state can be extended in the presence of conical defects for general SdS black holes.

\subsection*{Stretched horizon holography}
Let us reiterate that our protocol does not employ a particular realization of dS space holography. However, stretched horizon holography \cite{Susskind:2021omt} might allow for different modifications. In this framework, the stretched horizon represents a spacetime surface very close to the cosmological horizon where the holographic dual to dS space is located. It has been conjectured that the doubled-scaled SYK model might display some features of dS JT gravity \cite{Susskind:2022bia,Rahman:2022jsf,Goel:2023svz} at infinite temperatures, and it might be located at its stretched horizon. This approach mirrors several aspects of the AdS/CFT correspondence in that observables can be gravitationally dressed with respect to this surface.

\begin{figure}[t!]
    \centering
    \includegraphics[width=0.65\textwidth]{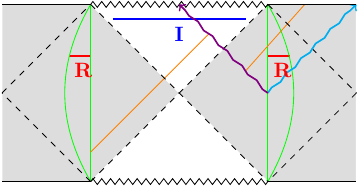}
    \caption{Alternative realization of the protocol employing the stretched horizon holographic proposal of \cite{Susskind:2021dfc}.}
    \label{fig:stretched horizon}
\end{figure}

It might be possible to perform the protocol in this framework following the same steps as illustrated in the protocol above according to the initial state preparation in empty dS space\footnote{Notice that the worldline observers in pure dS space, Fig. \ref{fig:dS plain}, would also have access to an island near the cosmological horizon \cite{Shaghoulian:2021cef}, which has been noticed to have different puzzling features, such as it moving backward in time \cite{Sybesma:2020fxg} and violating entanglement wedge nesting \cite{Shaghoulian:2021cef}.} and transmitting information {in} SdS black hole {backgrounds}. For probing the island region, we might consider an alternative configuration {with respect} to Fig. \ref{fig:Setup_Islands_detectors}, which we illustrate in Fig. \ref{fig:stretched horizon}. In this case, the pair of detectors are no longer located at $\mathcal{I}^+$, but rather on the stretched horizon, extending to the static patch observers. We expect the qualitative results would be just as the found for the configuration in Fig. \ref{fig:Setup_Islands_detectors} for recovering information about the island region.
It would also be interesting to study if a realization of the protocol could be performed explicitly in a double-scaled SYK model in the limit that might reproduce dS JT gravity \cite{Susskind:2022bia}. There have been different realizations of the GJW protocol in different contexts of the SYK \cite{Maldacena:2017axo,Maldacena:2018lmt,Gao:2019nyj,Gao:2023gta} that might be useful to {pursuit} this direction.

\subsection*{Traversable wormhole production rates}
Recently, \cite{Bintanja:2023vel} provided the false vacuum decay rate of a pair of Reissner-Nordstrom (RN) AdS$_4$ black holes into traversable wormholes connecting the geometries. In the microcanonical ensemble, they find that
\begin{equation}\label{eq:microcanonical ensemble rate}
    \Gamma\sim \rme^{\Delta S+2\beta(M_{\rm b}-M_0)},
\end{equation}
where $\Delta S$ is the total entropy of the pair of RN AdS$_4$ black holes, $\beta$ refers to their (inverse) temperature, $M_{\rm b}$ the black hole mass, and $M_0$ the mass of the traversable wormhole. It would be very interesting to study this type of process in the context of our protocol. Given that we required opposite energy pulses to produce the wormhole, we expect the contribution of the traversable wormhole from the black hole and the cosmological horizon patches will approximately cancel each other in the near-Nariai configuration, such that (\ref{eq:microcanonical ensemble rate}) might be modified to
\begin{equation}
    \Gamma\sim \rme^{\Delta S+2\beta_{\rm N}M_{\rm N}}~,
\end{equation}
according to the static patch observers. {We expect this result since} thermal fluctuations can also open traversable wormholes between the black hole and inflating regions for Nariai spacetimes. However, the size of such wormholes would not be significant enough to be of practical use. It might be useful to study the quasi-local thermodynamics approach to cosmological and black hole horizons \cite{Svesko:2022txo,Banihashemi:2022jys,Banihashemi:2022htw,Jacobson:2022gmo} to make this relation precise.

\section*{Acknowledgements}
We thank Lars Aalsma, Stefano Baiguera, Paul Balavoine, {Stefan Eccles}, Ben Freivogel, Viktor Jahnke, Jan Pieter van der Schaar, Andrea Legramandi, Raghu Mahajan, Xiao-Liang Qi, Aron Wall, Zhenbin Yang, and Ying Zhao for useful discussions, and Justin Charles Foster Sutherland for partially proofreading the manuscript. SEAG thanks the IFT-UAM/CSIC, the University of Amsterdam, the Delta Institute for Theoretical Physics, and the International Centre for Theoretical Physics for their hospitality and financial support during several phases of the project, and the Research Foundation - Flanders (FWO) for also providing mobility support (Grant No. K250423N). The work of SEAG is partially supported by the FWO Research Project G0H9318N and the inter-university project iBOF/21/084. EM is supported by Ama Mundu Technologies (Adoro te Devote Grant 2019). RE is supported by the Dushi Zhuanxiang Fellowship and acknowledges a Shuimu Scholarship as part of the Shuimu Tsinghua Scholar Program.

\appendix

\section{Hayden-Preskill protocol for near-Nariai black holes}\label{App:HaydenPreskill}
We refer the reader to \cite{Aalsma:2021kle} for an introduction to Hayden-Preskill protocol and information recovery in pure dS space, as well as the original publication \cite{Hayden:2007cs} in the black hole context.

\begin{figure}
\centering
\includegraphics[width=0.55\textwidth]{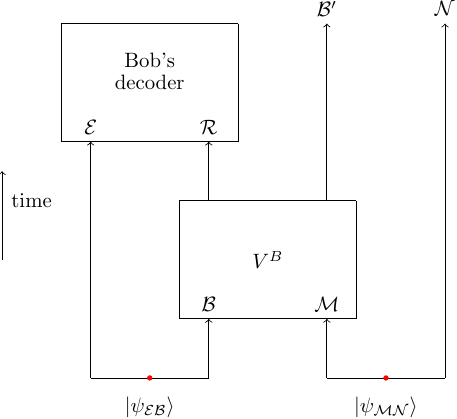}
    \caption{Quantum circuit representation of the Hayden-Preskill protocol, where $\ket{\psi_{\rm EB}}$ and $\ket{\psi_{\rm MN}}$ are maximally entangled states. Figure based on \cite{Hayden:2007cs}.}
    \label{fig:hp}
\end{figure}

The basic principle of operation of this protocol is illustrated in Fig. \ref{fig:hp}. 
From purely quantum information arguments, we search for a criterion establishing if the information transfer between the antipodal observers in the near-Nariai black hole spacetime {can be} successful.

We consider Fig. \ref{fig:hp} where one of the observers, Alice, possesses a subsystem $\mathcal{M}$ which is maximally entangled with a subsystem $\mathcal{N}$ carried on by Charlie. Meanwhile, an observer Bob has access to subsystem $\mathcal{E}$ (representing Hawking radiation) which is entangled with a black hole, $\mathcal{B}$. Alice waits until $\mathcal{E}$ is maximally entangled with $\mathcal{B}$ when she throws $\mathcal{M}$ into $\mathcal{B}$. We model the absorption of the message as a random unitary process $V^B$. This results in new subsystems, $\mathcal{B}'$ describing the black hole after absorption, and $\mathcal{R}$ for the radiation. In that process, $\mathcal{M}$, the purifier of $\mathcal{N}$, gets transferred from the subsystem $\mathcal{B}'$
\begin{equation}
    \ket{\psi}=\frac{1}{\sqrt{\norm{\mathcal{B}'}}}\sum_i\ket{i}_{\mathcal{B}'}\ket{0}_{\mathcal{R}}\ket{i}_{\mathcal{N}}~,
\end{equation}
to the subsystem $\mathcal{R}$
\begin{equation}
   \ket{\psi'}=\frac{1}{\sqrt{\norm{\mathcal{B}'}}}\sum_i\ket{0}_{\mathcal{B}'}\ket{i}_{\mathcal{R}}\ket{i}_{\mathcal{N}}~,
\end{equation}
where $\norm{B'}$ refers to the Hilbert space dimension of $\mathcal{B}'$. The criterion in \cite{Hayden:2007cs} for successful message transmission is that
\begin{equation}
    \norm{\rho_{\mathcal{MB}'}-\rho_{\mathcal{M}}\otimes\rho_{\mathcal{B}'}}\ll1~,
\end{equation}
    where $\rho_{\mathcal{M}}$, $\rho_{\mathcal{B}'}$, $\rho_{\mathcal{MB}'}$ refer to the reduced density matrices by tracing out the subsystems $\mathcal{B}'\mathcal{N}$, $\mathcal{NM}$ and $\mathcal{N}$. The norm is taken with respect to the Haar measure, defined as
\begin{equation}\label{eq:Haar measure}
\int \rmd V^B \norm{\rho_{\mathcal{MB}'}-\rho_{\mathcal{M}}\otimes\rho_{\mathcal{B}'}}=2^{-2C}~,
\end{equation}
with $C$ {denoting} the additional qubits in the radiation with respect to the number of qubits in Alice’s message. This means that as soon as the black hole has radiated more information than the {amount} encoded in the {Alice's original message}, Bob gains access to it.

We can directly apply this protocol to transmit light signals in the near-Nariai black hole background in Fig. \ref{fig:bath}. However, notice that in this case, Alice does not need to wait until the Page time to throw $\mathcal{M}$ into $\mathcal{B}$. {The reason is that} the preparation of the state from the BD vacuum already maximally entangles the pair of antipodal observers.
As we saw in Sec. \ref{sec:Set up}, we achieved information recovery by transmitting signals across the traversable wormhole produced in the near-Nariai black hole over a time scale determined {by} (\ref{eq:Time scales BH}). Meanwhile, if we perform the information transfer in the inflationary region, the time scale for recovery is given by (\ref{eq:time scale CH}).

The reader is referred to \cite{Hayden:2007cs,Aalsma:2021kle} for considerations about how to avoid the cloning paradox in HP protocol for the black holes and in pure dS space.

\section{S-wave reduction of SdS black holes}\label{App:JT from SdS}
We follow the derivation of the full-reduction dS JT gravity model given \cite{Svesko:2022txo}. In our notation, hats are used for $(d+1)$-dimensional quantities and unhatted for $(1+1)$-dimensional ones. Starting from Einstein gravity in $(d+1)$-dimensions
\begin{equation}\label{eq:SdS action}
    I=\frac{1}{16\pi G_N}\int_{\hat{\mathscr{M}}}\rmd^{d+1}X\sqrt{-\hat g}(\hat R-2\Lambda)+\frac{1}{8\pi G_N}\int_{\partial\hat{\mathscr{M}}}\rmd^{d}Y\sqrt{-\hat h}\,\hat K~,
\end{equation}
with a positive cosmological constant
\begin{equation}\label{eq:Lambda C.C.}
    \Lambda=\frac{d(d-1)}{2\ell^2}>0~.
\end{equation}
The general solution is given by the SdS metric (\ref{metric1}). We consider dimensional reduction with the metric ansatz,
\begin{equation}
\begin{aligned}
    \rmd s^2&=g_{\mu\nu}(x)\rmd x^\mu\rmd x^\nu+r_{\rm N}^2\Phi^{\frac{2}{d-1}}\rmd\Omega^2_{d-1}
\end{aligned}
\end{equation}
where $\mu,\, \nu = 0,\, 1,$ and $\Phi(x)$ is the dilaton. Next, we perform a Weyl rescaling $g_{\mu\nu}\rightarrow \frac{1}{\sqrt{d}}\Phi^{-\frac{d-2}{2(d-1)}}g_{\mu\nu}$ to express the EFT coming from the spherical reduction of SdS black holes as:
\begin{equation}\label{eq:Exact dimensional reduction}
    I=\frac{1}{16\pi G_2}\int_{\mathscr{M}}\rmd^2x\sqrt{-g}\qty[\Phi R+U(\Phi)]+\frac{1}{8\pi G_2}\int_{\partial\mathscr{M}}\sqrt{-h}\Phi K,
\end{equation}
with
\begin{equation}\label{eq:U(Phi)}
    \begin{aligned}
        U(\Phi)&=\frac{d-1}{\ell^2}\qty(\Phi^{-\frac{1}{d-1}}-\Phi^{\frac{1}{d-1}})~,\\
    G_2&=\frac{G_N}{\ell^{d-1}\Omega_{d-1}}~.
    \end{aligned}
\end{equation}
The resulting equations of motion are
    \begin{align}\label{eq:EOM Dilaton Grav}
&(\nabla_\mu\nabla_\nu-g_{\mu\nu}\nabla^2)\Phi+\frac{g_{\mu\nu}}{2}U(\Phi)=-8\pi G_2T_{\mu\nu}^{(m)}~,\\
    &R=U'(\Phi)~,\label{eq:R dilaton}
\end{align}
where we added a $(1+1)$-dimensional bulk-matter term $I_m$ generating the matter stress-tensor $T_{\mu\nu}^{\rm (m)}$. We also notice that the minimum of the potential (\ref{eq:U(Phi)}) is located at $\Phi=1$, corresponding to the Nariai black hole limit, for which
\begin{equation}
\begin{aligned}
    U(\Phi=1)&=0~,\\
    \eval{\dv{U}{\Phi}}_{\Phi=1}&=-\frac{2}{\ell^2}~.
\end{aligned}   
\end{equation}
If we take $\Phi=\phi_0+\phi$, with $\phi_0=1$, we can perform a series expansion for $\phi\ll1$ and truncate the action (\ref{eq:Exact dimensional reduction}) to first order in $\phi$ we recover the so-called dS$_2$ JT full-reduction model with the minimally coupled bulk matter fields that we introduced in (\ref{eq:EOM Dilaton Grav})
\begin{equation}\label{eq:I JT}
    I_{\rm dS\,JT}=I_0+\frac{1}{16\pi G_2}\int_{\mathscr{M}}\rmd^2x\sqrt{-{g}}\qty(\phi{R}-\frac{2}{\ell^2}\phi)+\frac{1}{8\pi G_2}\int_{\partial\mathscr{M}}\rmd y\sqrt{-{h}}\,\phi {K}+I_{\rm matter}~,
\end{equation}
where $I_0=\frac{\phi_0}{16\pi G_2}\int_{\mathscr{M}}\rmd^2x\sqrt{-{g}}R$ refers to the topological term in the action, and $I_{\rm matter}$ is the matter content. Notice that the addition of the dilaton mode breaks asymptotic dS$_2$ space isometries, namely reparametrizations of coordinates at $\mathcal{I}^+$ \cite{Maldacena:2019cbz}.

\section{Shifts from metric gluing}\label{App:Shift SdS}
In {a} seminal work {by Shenker and Stanford} \cite{Shenker:2013pqa}, a convenient way to obtain the shift in trajectory experienced by a probe message following a null-like geodesic was stated. We extend this formalism to SdS space. Let us express the Kruskal coordinates in (\ref{eq:Metric Kruskal}) in terms of static patch coordinates (\ref{metric1}, \ref{eq:blackeningfactor}):
\begin{equation}
    \begin{aligned}
U_{\rm b,\,c}&=\rme^{\frac{f'(r_{\rm b,\,c})}{2}\left(r_{*}(r)-t\right)}~,\\
V_{\rm b,c}&=-\rme^{\frac{f'(r_{\rm b,\,c})}{2}\left(r_{*}(r)+t\right)}\,,
\end{aligned}
\end{equation}
centered at the horizons $r_{\rm b,\, \rm c}$ for $0\leq r< r_{\mathcal{O}}$ and $r_{\mathcal{O}} \leq r < \infty$ respectively. Evaluating the tortoise coordinate $\rmd r_{*}=\rmd r/f(r)$ results in 
\begin{equation}
    \rmd s^2=\left\{
    \begin{array}{ll}
        -\frac{4 f(r)}{f'(r_{\rm b})}\rme^{-f'(r_{\rm b})r_{*}(r)}\rmd U_{\rm b}\,\rmd V_{\rm b} +r^2 \rmd \Omega_{d-1}^2\equiv \rmd s^2_{\rm b}& \quad \, 0\leq r< r_{\mathcal{O}} \\
         -\frac{4 f(r)}{f'(r_{\rm c})}\rme^{-f'(r_{\rm c})r_{*}(r)}\rmd U_{\rm c}\rmd V_{\rm c} +r^2 \rmd \Omega_{d-1}^2 \equiv \rmd s^2_{\rm c}& \quad \,  r_{\mathcal{O}} \leq r < \infty
    \end{array}
\right.\,.
\end{equation}
When gluing two SdS space-times of different masses to each other, both the black hole and cosmological line elements will experience shifts along $U=U_0$ and $V=V_0$ respectively,
\begin{equation}
    \begin{aligned}
    \rmd s^2_{\rm b}&=
    \begin{cases}
         \rmd s^2_{\rm b}& \quad \, U\leq U_0\\
        \rmd \tilde{s}^2_{\rm b}& \quad \,  U_0< U
    \end{cases}~,\\
\rmd s^2_{\rm c}&=
    \begin{cases}
         \rmd s^2_{\rm c}& \quad \, V\leq V_0\\
        \rmd \tilde{s}^2_{\rm c}& \quad \,  V_0< V
    \end{cases}~,
\end{aligned}
\end{equation}
where we simplified the notation to $(U_{\rm b,\rm c},\,V_{\rm b,\rm c})\rightarrow (U,\,V)$ for more clarity. We focus on the black hole region $0\leq r< r_{\mathcal{O}}$, and introduce a perturbation at $t=t_0$ forming a shell propagating along the surface
\begin{equation}
    \begin{aligned}
        \tilde{U}_0&=\rme^{\frac{\tilde{f}'(\tilde{r_{\rm b}})}{2}\left(\tilde{r}_{*}(r_{\mathcal{O}})-t_0\right)}~,\\
    U_0&=\rme^{\frac{f'(r_{\rm b})}{2}\left(r_{*}(r_{\mathcal{O}})-t_0\right)}\,,
    \end{aligned}
\end{equation}
such that the gluing/matching condition relating $\tilde{V}$ to $V$ can be expressed as follows
\begin{equation}
\begin{aligned}
    \tilde{U}_0 \tilde{V}&=-\rme^{\tilde{f}'(\tilde{r}_{\rm b})\tilde{r}_{*}(r)}~,\\
    \quad U_0 V&=-\rme^{f'(r_{\rm b})r_{*}(r)}\,.
\end{aligned}
\end{equation}
When working in the small energy limit $E/M=\epsilon\ll1$, $\tilde{U}_0$ can be well approximated by $U_0$, allowing us to deduce that, at first order, the shift along the $V$ coordinate; {which} can be expressed
\begin{equation}
\begin{aligned}
    \tilde{V}&=V-\frac{\epsilon}{U_o}\left(\frac{\partial}{\partial \epsilon}\rme^{\tilde{f}'(\tilde{r}_{\rm b})\tilde{r}_{*}(r)}\right)_{\epsilon=0}\\
    &\equiv V+\alpha\,.
\end{aligned}
\end{equation}
As was pointed out in \cite{Shenker:2013pqa}, $\rme^{f'(r_{\rm b}) r_{*}(r)}$ can be compactly written (in $d=3$) as 
\begin{equation}
\begin{aligned}
   \rme^{f'(r_{\rm b}) r_{*}(r)}&=\left(\frac{r-r_{\rm b}}{\ell}\right)\left(\frac{r-r_{\rm c}}{\ell }\right){}^{-\frac{r_{\rm c} \left(2 r_{\rm b}+r_{\rm c}\right)}{r_{\rm b} \left(r_{\rm b}+2 r_{\rm c}\right)}}
   \left(\frac{r_{\rm b}+r_{\rm c}+r}{\ell }\right){}^{\frac{r_{\rm c}^2-r_{\rm b}^2}{r_{\rm b} \left(r_{\rm b}+2 r_{\rm c}\right)}}\\
   &\equiv (r-r_{\rm b})\,C_{\rm b}(r,r_{\rm b})\,,
\end{aligned}
\end{equation}
and the shift $\alpha$ becomes
\begin{equation}
\begin{aligned}
    \alpha&=\frac{G_N E \beta_{\rm b}}{ 2\pi r_{\rm b} U_0}\frac{\partial}{\partial r_{\rm b}}\bigg((r_{\rm b}-r)\,C_{\rm b}(r,r_{\rm b})\bigg)\,,
\end{aligned}
\end{equation}
 where we used $\tilde{M}=M(1-\epsilon)$ and {we} normalized the temperature with respect of the observer radius\footnote{Using the normalized temperature in $dM= T dS$ allows to write \begin{equation}
     \frac{\partial r_{\rm h}}{\partial M}=\frac{\partial r_{\rm h}}{\partial S_{\rm h}}\frac{\partial S_{\rm h}}{\partial M}=\frac{\beta_{\rm h}G_N}{2\pi r_{\rm b}}\,.
 \end{equation}
 }. Obviously, the same is happening in the cosmological region $r_{\mathcal{O}} \leq r < \infty$ (along $V$). {Thus,} the coordinate shifts can be expressed as
 \begin{equation}
 \begin{aligned}
     \alpha_{\rm b}&=\frac{r_{\rm b}}{2 U_0}\frac{s_{\rm b}}{S_{\rm b}}\frac{\partial}{\partial r_{\rm b}}\bigg((r_{\rm b}-r)\,C_{\rm b}(r,r_{\rm b})\bigg)~,\\
     \alpha_{\rm c}&=\frac{r_{\rm c}}{2 V_0}\frac{s_{\rm c}}{S_{\rm c}}\frac{\partial}{\partial r_{\rm c}}\bigg((r_{\rm c}-r)\,C_{\rm c}(r,r_{\rm c})\bigg)\,,
 \end{aligned}
 \end{equation}
where we restored the indices 
 and introduced the wormhole entropies $s_{\rm b, \rm c}=E\beta_{\rm b, \rm c}$. Notably, in the Nariai limit $r_{\rm b,\rm c}\rightarrow r_{\rm N}$ the function $C_{\rm N}(r,r_{\rm N})=(r-r_N)^{-1}$. This means that the shifts tend to vanish. However, considering $M\neq M_N$ and inserting the shockwave far in the past, the value for $r$ is getting pushed towards the respective horizon radius. We then find
\begin{equation}
\begin{aligned}
    \alpha_{\rm b}&=\frac{r_{\rm b}}{2U_0}\frac{s_{\rm b}}{S_{\rm b}}C_{\rm b}(r_{\rm b},r_{\rm b})~,\\
    \alpha_{\rm c}&=\frac{r_{\rm c}}{2V_0}\frac{s_{\rm c}}{S_{\rm c}}C_{\rm c}(r_{\rm c},r_{\rm c})\,.
\end{aligned}
\end{equation}
Notice, again, that a probe message sent towards the black hole will experience a different shift than a probe message sent towards the cosmological horizon when encountering shockwaves of the same magnitude of energy, $E$.

\bibliographystyle{JHEP}
\bibliography{references.bib}

\providecommand{\href}[2]{#2}\begingroup\raggedright\begin{thebibliography}{10}

\bibitem{Kundu:2021nwp}
A.~Kundu, \emph{{Wormholes and holography: an introduction}}, \href{https://doi.org/10.1140/epjc/s10052-022-10376-z}{\emph{Eur. Phys. J. C} {\bfseries 82} (2022) 447} [\href{https://arxiv.org/abs/2110.14958}{{\ttfamily 2110.14958}}].

\bibitem{Maldacena:2013xja}
J.~Maldacena and L.~Susskind, \emph{{Cool horizons for entangled black holes}}, \href{https://doi.org/10.1002/prop.201300020}{\emph{Fortsch. Phys.} {\bfseries 61} (2013) 781} [\href{https://arxiv.org/abs/1306.0533}{{\ttfamily 1306.0533}}].

\bibitem{Freivogel:2019lej}
B.~Freivogel, V.~Godet, E.~Morvan, J.F.~Pedraza and A.~Rotundo, \emph{{Lessons on eternal traversable wormholes in AdS}}, \href{https://doi.org/10.1007/JHEP07(2019)122}{\emph{JHEP} {\bfseries 07} (2019) 122} [\href{https://arxiv.org/abs/1903.05732}{{\ttfamily 1903.05732}}].

\bibitem{Morris:1988tu}
M.S.~Morris, K.S.~Thorne and U.~Yurtsever, \emph{{Wormholes, Time Machines, and the Weak Energy Condition}}, \href{https://doi.org/10.1103/PhysRevLett.61.1446}{\emph{Phys. Rev. Lett.} {\bfseries 61} (1988) 1446}.

\bibitem{Friedman:1993ty}
J.L.~Friedman, K.~Schleich and D.M.~Witt, \emph{{Topological censorship}}, \href{https://doi.org/10.1103/PhysRevLett.71.1486}{\emph{Phys. Rev. Lett.} {\bfseries 71} (1993) 1486} [\href{https://arxiv.org/abs/gr-qc/9305017}{{\ttfamily gr-qc/9305017}}].

\bibitem{Wall:2009wi}
A.C.~Wall, \emph{{Proving the Achronal Averaged Null Energy Condition from the Generalized Second Law}}, \href{https://doi.org/10.1103/PhysRevD.81.024038}{\emph{Phys. Rev. D} {\bfseries 81} (2010) 024038} [\href{https://arxiv.org/abs/0910.5751}{{\ttfamily 0910.5751}}].

\bibitem{Visser:1996iw}
M.~Visser, \emph{{Gravitational vacuum polarization. 1: Energy conditions in the Hartle-Hawking vacuum}}, \href{https://doi.org/10.1103/PhysRevD.54.5103}{\emph{Phys. Rev. D} {\bfseries 54} (1996) 5103} [\href{https://arxiv.org/abs/gr-qc/9604007}{{\ttfamily gr-qc/9604007}}].

\bibitem{Visser:1995cc}
M.~Visser, \emph{{Lorentzian wormholes: From Einstein to Hawking}}, American Institute of Physics Press (1995).

\bibitem{Gao:2016bin}
P.~Gao, D.L.~Jafferis and A.C.~Wall, \emph{{Traversable Wormholes via a Double Trace Deformation}}, \href{https://doi.org/10.1007/JHEP12(2017)151}{\emph{JHEP} {\bfseries 12} (2017) 151} [\href{https://arxiv.org/abs/1608.05687}{{\ttfamily 1608.05687}}].

\bibitem{Maldacena:2017axo}
J.~Maldacena, D.~Stanford and Z.~Yang, \emph{{Diving into traversable wormholes}}, \href{https://doi.org/10.1002/prop.201700034}{\emph{Fortsch. Phys.} {\bfseries 65} (2017) 1700034} [\href{https://arxiv.org/abs/1704.05333}{{\ttfamily 1704.05333}}].

\bibitem{Maldacena:2018lmt}
J.~Maldacena and X.-L.~Qi, \emph{{Eternal traversable wormhole}},  \href{https://arxiv.org/abs/1804.00491}{{\ttfamily 1804.00491}}.

\bibitem{Maldacena:2018gjk}
J.~Maldacena, A.~Milekhin and F.~Popov, \emph{{Traversable wormholes in four dimensions}},  \href{https://arxiv.org/abs/1807.04726}{{\ttfamily 1807.04726}}.

\bibitem{Maldacena:2020sxe}
J.~Maldacena and A.~Milekhin, \emph{{Humanly traversable wormholes}}, \href{https://doi.org/10.1103/PhysRevD.103.066007}{\emph{Phys. Rev. D} {\bfseries 103} (2021) 066007} [\href{https://arxiv.org/abs/2008.06618}{{\ttfamily 2008.06618}}].

\bibitem{Bintanja:2021xfs}
S.~Bintanja, R.~Esp\'\i{}ndola, B.~Freivogel and D.~Nikolakopoulou, \emph{{How to make traversable wormholes: eternal AdS$_{4}$ wormholes from coupled CFT\textquoteright{}s}}, \href{https://doi.org/10.1007/JHEP10(2021)173}{\emph{JHEP} {\bfseries 10} (2021) 173} [\href{https://arxiv.org/abs/2102.06628}{{\ttfamily 2102.06628}}].

\bibitem{deBoer:2018ibj}
J.~de~Boer, R.~Van~Breukelen, S.F.~Lokhande, K.~Papadodimas and E.~Verlinde, \emph{{On the interior geometry of a typical black hole microstate}}, \href{https://doi.org/10.1007/JHEP05(2019)010}{\emph{JHEP} {\bfseries 05} (2019) 010} [\href{https://arxiv.org/abs/1804.10580}{{\ttfamily 1804.10580}}].

\bibitem{DeBoer:2019yoe}
J.~De~Boer, R.~Van~Breukelen, S.F.~Lokhande, K.~Papadodimas and E.~Verlinde, \emph{{Probing typical black hole microstates}}, \href{https://doi.org/10.1007/JHEP01(2020)062}{\emph{JHEP} {\bfseries 01} (2020) 062} [\href{https://arxiv.org/abs/1901.08527}{{\ttfamily 1901.08527}}].

\bibitem{Hartman:2020khs}
T.~Hartman, Y.~Jiang and E.~Shaghoulian, \emph{{Islands in cosmology}}, \href{https://doi.org/10.1007/JHEP11(2020)111}{\emph{JHEP} {\bfseries 11} (2020) 111} [\href{https://arxiv.org/abs/2008.01022}{{\ttfamily 2008.01022}}].

\bibitem{Aguilar-Gutierrez:2021bns}
S.E.~Aguilar-Gutierrez, A.~Chatwin-Davies, T.~Hertog, N.~Pinzani-Fokeeva and B.~Robinson, \emph{{Islands in Multiverse Models}}, \href{https://doi.org/10.1007/JHEP11(2021)212}{\emph{JHEP} {\bfseries 11} (2021) 212} [\href{https://arxiv.org/abs/2108.01278}{{\ttfamily 2108.01278}}].

\bibitem{Bousso:2022gth}
R.~Bousso and E.~Wildenhain, \emph{{Islands in closed and open universes}}, \href{https://doi.org/10.1103/PhysRevD.105.086012}{\emph{Phys. Rev. D} {\bfseries 105} (2022) 086012} [\href{https://arxiv.org/abs/2202.05278}{{\ttfamily 2202.05278}}].

\bibitem{Espindola:2022fqb}
R.~Esp\'\i{}ndola, B.~Najian and D.~Nikolakopoulou, \emph{{Islands in FRW Cosmologies}},  \href{https://arxiv.org/abs/2203.04433}{{\ttfamily 2203.04433}}.

\bibitem{Ben-Dayan:2022nmb}
I.~Ben-Dayan, M.~Hadad and E.~Wildenhain, \emph{{Islands in the fluid: islands are common in cosmology}}, \href{https://doi.org/10.1007/JHEP03(2023)077}{\emph{JHEP} {\bfseries 03} (2023) 077} [\href{https://arxiv.org/abs/2211.16600}{{\ttfamily 2211.16600}}].

\bibitem{Aalsma:2021kle}
L.~Aalsma, A.~Cole, E.~Morvan, J.P.~van~der Schaar and G.~Shiu, \emph{{Shocks and information exchange in de Sitter space}}, \href{https://doi.org/10.1007/JHEP10(2021)104}{\emph{JHEP} {\bfseries 10} (2021) 104} [\href{https://arxiv.org/abs/2105.12737}{{\ttfamily 2105.12737}}].

\bibitem{Gao:2000ga}
S.~Gao and R.M.~Wald, \emph{{Theorems on gravitational time delay and related issues}}, \href{https://doi.org/10.1088/0264-9381/17/24/305}{\emph{Class. Quant. Grav.} {\bfseries 17} (2000) 4999} [\href{https://arxiv.org/abs/gr-qc/0007021}{{\ttfamily gr-qc/0007021}}].

\bibitem{Dyson:1960xib}
F.J.~Dyson, \emph{{Search for Artificial Stellar Sources of Infrared Radiation}}, \href{https://doi.org/10.1126/science.131.3414.1667}{\emph{Science} {\bfseries 131} (1960) 1667}.

\bibitem{Freivogel:2019whb}
B.~Freivogel, D.A.~Galante, D.~Nikolakopoulou and A.~Rotundo, \emph{{Traversable wormholes in AdS and bounds on information transfer}}, \href{https://doi.org/10.1007/JHEP01(2020)050}{\emph{JHEP} {\bfseries 01} (2020) 050} [\href{https://arxiv.org/abs/1907.13140}{{\ttfamily 1907.13140}}].

\bibitem{Ginsparg:1982rs}
P.H.~Ginsparg and M.J.~Perry, \emph{{Semiclassical Perdurance of de Sitter Space}}, \href{https://doi.org/10.1016/0550-3213(83)90636-3}{\emph{Nucl. Phys. B} {\bfseries 222} (1983) 245}.

\bibitem{Bousso:1996au}
R.~Bousso and S.W.~Hawking, \emph{{Pair creation of black holes during inflation}}, \href{https://doi.org/10.1103/PhysRevD.54.6312}{\emph{Phys. Rev. D} {\bfseries 54} (1996) 6312} [\href{https://arxiv.org/abs/gr-qc/9606052}{{\ttfamily gr-qc/9606052}}].

\bibitem{Draper:2022xzl}
P.~Draper and S.~Farkas, \emph{{de Sitter Black Holes as Constrained States in the Euclidean Path Integral}},  \href{https://arxiv.org/abs/2203.02426}{{\ttfamily 2203.02426}}.

\bibitem{Morvan:2022ybp}
E.K.~Morvan, J.P.~van~der Schaar and M.R.~Visser, \emph{{On the Euclidean action of de Sitter black holes and constrained instantons}}, \href{https://doi.org/10.21468/SciPostPhys.14.2.022}{\emph{SciPost Phys.} {\bfseries 14} (2023) 022} [\href{https://arxiv.org/abs/2203.06155}{{\ttfamily 2203.06155}}].

\bibitem{Morvan:2022aon}
E.K.~Morvan, J.P.~van~der Schaar and M.R.~Visser, \emph{{Action, entropy and pair creation rate of charged black holes in de Sitter space}},  \href{https://arxiv.org/abs/2212.12713}{{\ttfamily 2212.12713}}.

\bibitem{Witten:2001kn}
E.~Witten, \emph{{Quantum gravity in de Sitter space}},  in \emph{{Strings 2001: International Conference}}, 6, 2001 [\href{https://arxiv.org/abs/hep-th/0106109}{{\ttfamily hep-th/0106109}}].

\bibitem{Strominger:2001pn}
A.~Strominger, \emph{{The dS / CFT correspondence}}, \href{https://doi.org/10.1088/1126-6708/2001/10/034}{\emph{JHEP} {\bfseries 10} (2001) 034} [\href{https://arxiv.org/abs/hep-th/0106113}{{\ttfamily hep-th/0106113}}].

\bibitem{Maldacena:2002vr}
J.M.~Maldacena, \emph{{Non-Gaussian features of primordial fluctuations in single field inflationary models}}, \href{https://doi.org/10.1088/1126-6708/2003/05/013}{\emph{JHEP} {\bfseries 05} (2003) 013} [\href{https://arxiv.org/abs/astro-ph/0210603}{{\ttfamily astro-ph/0210603}}].

\bibitem{Galante:2023uyf}
D.A.~Galante, \emph{{Modave lectures on de Sitter space \& holography}}, \href{https://doi.org/10.22323/1.435.0003}{\emph{PoS} {\bfseries Modave2022} (2023) 003} [\href{https://arxiv.org/abs/2306.10141}{{\ttfamily 2306.10141}}].

\bibitem{Penington:2019npb}
G.~Penington, \emph{{Entanglement Wedge Reconstruction and the Information Paradox}}, \href{https://doi.org/10.1007/JHEP09(2020)002}{\emph{JHEP} {\bfseries 09} (2020) 002} [\href{https://arxiv.org/abs/1905.08255}{{\ttfamily 1905.08255}}].

\bibitem{Almheiri:2019psf}
A.~Almheiri, N.~Engelhardt, D.~Marolf and H.~Maxfield, \emph{{The entropy of bulk quantum fields and the entanglement wedge of an evaporating black hole}}, \href{https://doi.org/10.1007/JHEP12(2019)063}{\emph{JHEP} {\bfseries 12} (2019) 063} [\href{https://arxiv.org/abs/1905.08762}{{\ttfamily 1905.08762}}].

\bibitem{Almheiri:2019hni}
A.~Almheiri, R.~Mahajan, J.~Maldacena and Y.~Zhao, \emph{{The Page curve of Hawking radiation from semiclassical geometry}}, \href{https://doi.org/10.1007/JHEP03(2020)149}{\emph{JHEP} {\bfseries 03} (2020) 149} [\href{https://arxiv.org/abs/1908.10996}{{\ttfamily 1908.10996}}].

\bibitem{Chen:2019gbt}
C.-F.~Chen, G.~Penington and G.~Salton, \emph{{Entanglement Wedge Reconstruction using the Petz Map}}, \href{https://doi.org/10.1007/JHEP01(2020)168}{\emph{JHEP} {\bfseries 01} (2020) 168} [\href{https://arxiv.org/abs/1902.02844}{{\ttfamily 1902.02844}}].

\bibitem{Penington:2019kki}
G.~Penington, S.H.~Shenker, D.~Stanford and Z.~Yang, \emph{{Replica wormholes and the black hole interior}}, \href{https://doi.org/10.1007/JHEP03(2022)205}{\emph{JHEP} {\bfseries 03} (2022) 205} [\href{https://arxiv.org/abs/1911.11977}{{\ttfamily 1911.11977}}].

\bibitem{Zhao:2020wgp}
Y.~Zhao, \emph{{Petz map and Python\textquoteright{}s lunch}}, \href{https://doi.org/10.1007/JHEP11(2020)038}{\emph{JHEP} {\bfseries 11} (2020) 038} [\href{https://arxiv.org/abs/2003.03406}{{\ttfamily 2003.03406}}].

\bibitem{Bak:2021qbo}
D.~Bak, C.~Kim, S.-H.~Yi and J.~Yoon, \emph{{Python\textquoteright{}s lunches in Jackiw-Teitelboim gravity with matter}}, \href{https://doi.org/10.1007/JHEP04(2022)175}{\emph{JHEP} {\bfseries 04} (2022) 175} [\href{https://arxiv.org/abs/2112.04224}{{\ttfamily 2112.04224}}].

\bibitem{Vardian:2023fce}
N.~Vardian, \emph{{Black hole interior Petz map reconstruction and Papadodimas-Raju proposal}},  \href{https://arxiv.org/abs/2307.01858}{{\ttfamily 2307.01858}}.

\bibitem{Bahiru:2022ukn}
E.~Bahiru and N.~Vardian, \emph{{Explicit reconstruction of the entanglement wedge via the Petz map}}, \href{https://doi.org/10.1007/JHEP07(2023)025}{\emph{JHEP} {\bfseries 07} (2023) 025} [\href{https://arxiv.org/abs/2210.00602}{{\ttfamily 2210.00602}}].

\bibitem{Geng:2020kxh}
H.~Geng, \emph{{Non-local entanglement and fast scrambling in de-Sitter holography}}, \href{https://doi.org/10.1016/j.aop.2021.168402}{\emph{Annals Phys.} {\bfseries 426} (2021) 168402} [\href{https://arxiv.org/abs/2005.00021}{{\ttfamily 2005.00021}}].

\bibitem{Geng:2021wcq}
H.~Geng, Y.~Nomura and H.-Y.~Sun, \emph{{An Information Paradox and Its Resolution in de Sitter Holography}},  \href{https://arxiv.org/abs/2103.07477}{{\ttfamily 2103.07477}}.

\bibitem{Bousso:2022hlz}
R.~Bousso and G.~Penington, \emph{{Entanglement wedges for gravitating regions}}, \href{https://doi.org/10.1103/PhysRevD.107.086002}{\emph{Phys. Rev. D} {\bfseries 107} (2023) 086002} [\href{https://arxiv.org/abs/2208.04993}{{\ttfamily 2208.04993}}].

\bibitem{Bousso:2023sya}
R.~Bousso and G.~Penington, \emph{{Holograms In Our World}},  \href{https://arxiv.org/abs/2302.07892}{{\ttfamily 2302.07892}}.

\bibitem{Chen:2020tes}
Y.~Chen, V.~Gorbenko and J.~Maldacena, \emph{{Bra-ket wormholes in gravitationally prepared states}}, \href{https://doi.org/10.1007/JHEP02(2021)009}{\emph{JHEP} {\bfseries 02} (2021) 009} [\href{https://arxiv.org/abs/2007.16091}{{\ttfamily 2007.16091}}].

\bibitem{Piao:2023vgm}
Y.-S.~Piao, \emph{{Implication of the island rule for inflation and primordial perturbations}}, \href{https://doi.org/10.1103/PhysRevD.107.123509}{\emph{Phys. Rev. D} {\bfseries 107} (2023) 123509} [\href{https://arxiv.org/abs/2301.07403}{{\ttfamily 2301.07403}}].

\bibitem{Sybesma:2020fxg}
W.~Sybesma, \emph{{Pure de Sitter space and the island moving back in time}}, \href{https://doi.org/10.1088/1361-6382/abff9a}{\emph{Class. Quant. Grav.} {\bfseries 38} (2021) 145012} [\href{https://arxiv.org/abs/2008.07994}{{\ttfamily 2008.07994}}].

\bibitem{Goswami:2022ylc}
K.~Goswami and K.~Narayan, \emph{{Small Schwarzschild de Sitter black holes, quantum extremal surfaces and islands}}, \href{https://doi.org/10.1007/JHEP10(2022)031}{\emph{JHEP} {\bfseries 10} (2022) 031} [\href{https://arxiv.org/abs/2207.10724}{{\ttfamily 2207.10724}}].

\bibitem{Aalsma:2021bit}
L.~Aalsma and W.~Sybesma, \emph{{The Price of Curiosity: Information Recovery in de Sitter Space}},  \href{https://arxiv.org/abs/2104.00006}{{\ttfamily 2104.00006}}.

\bibitem{Levine:2022wos}
A.~Levine and E.~Shaghoulian, \emph{{Encoding beyond cosmological horizons in de Sitter JT gravity}}, \href{https://doi.org/10.1007/JHEP02(2023)179}{\emph{JHEP} {\bfseries 02} (2023) 179} [\href{https://arxiv.org/abs/2204.08503}{{\ttfamily 2204.08503}}].

\bibitem{Balasubramanian:2020xqf}
V.~Balasubramanian, A.~Kar and T.~Ugajin, \emph{{Islands in de Sitter space}}, \href{https://doi.org/10.1007/JHEP02(2021)072}{\emph{JHEP} {\bfseries 02} (2021) 072} [\href{https://arxiv.org/abs/2008.05275}{{\ttfamily 2008.05275}}].

\bibitem{Kames-King:2021etp}
J.~Kames-King, E.M.H.~Verheijden and E.P.~Verlinde, \emph{{No Page curves for the de Sitter horizon}}, \href{https://doi.org/10.1007/JHEP03(2022)040}{\emph{JHEP} {\bfseries 03} (2022) 040} [\href{https://arxiv.org/abs/2108.09318}{{\ttfamily 2108.09318}}].

\bibitem{Baek:2022ozg}
J.-H.~Baek and K.-S.~Choi, \emph{{Islands in proliferating de Sitter spaces}}, \href{https://doi.org/10.1007/JHEP05(2023)098}{\emph{JHEP} {\bfseries 05} (2023) 098} [\href{https://arxiv.org/abs/2212.14753}{{\ttfamily 2212.14753}}].

\bibitem{Aalsma:2022swk}
L.~Aalsma, S.E.~Aguilar-Gutierrez and W.~Sybesma, \emph{{An outsider\textquoteright{}s perspective on information recovery in de Sitter space}}, \href{https://doi.org/10.1007/JHEP01(2023)129}{\emph{JHEP} {\bfseries 01} (2023) 129} [\href{https://arxiv.org/abs/2210.12176}{{\ttfamily 2210.12176}}].

\bibitem{Teresi:2021qff}
D.~Teresi, \emph{{Islands and the de Sitter entropy bound}}, \href{https://doi.org/10.1007/JHEP10(2022)179}{\emph{JHEP} {\bfseries 10} (2022) 179} [\href{https://arxiv.org/abs/2112.03922}{{\ttfamily 2112.03922}}].

\bibitem{Jiang:2024xnd}
W.-H.~Jiang and Y.-S.~Piao, \emph{{Bounded islands in dS$_{2}$ multiverse model}},  \href{https://arxiv.org/abs/2403.18420}{{\ttfamily 2403.18420}}.

\bibitem{Chandrasekaran:2022cip}
V.~Chandrasekaran, R.~Longo, G.~Penington and E.~Witten, \emph{{An algebra of observables for de Sitter space}}, \href{https://doi.org/10.1007/JHEP02(2023)082}{\emph{JHEP} {\bfseries 02} (2023) 082} [\href{https://arxiv.org/abs/2206.10780}{{\ttfamily 2206.10780}}].

\bibitem{Witten:2023qsv}
E.~Witten, \emph{{Algebras, Regions, and Observers}},  \href{https://arxiv.org/abs/2303.02837}{{\ttfamily 2303.02837}}.

\bibitem{Gomez:2022eui}
C.~Gomez, \emph{{Cosmology as a Crossed Product}},  \href{https://arxiv.org/abs/2207.06704}{{\ttfamily 2207.06704}}.

\bibitem{Seo:2022pqj}
M.-S.~Seo, \emph{{von Neumann algebra description of inflationary cosmology}},  \href{https://arxiv.org/abs/2212.05637}{{\ttfamily 2212.05637}}.

\bibitem{Gomez:2023wrq}
C.~Gomez, \emph{{Entanglement, Observers and Cosmology: a view from von Neumann Algebras}},  \href{https://arxiv.org/abs/2302.14747}{{\ttfamily 2302.14747}}.

\bibitem{Gomez:2023upk}
C.~Gomez, \emph{{Clocks, Algebras and Cosmology}},  \href{https://arxiv.org/abs/2304.11845}{{\ttfamily 2304.11845}}.

\bibitem{Jensen:2023yxy}
K.~Jensen, J.~Sorce and A.J.~Speranza, \emph{{Generalized entropy for general subregions in quantum gravity}}, \href{https://doi.org/10.1007/JHEP12(2023)020}{\emph{JHEP} {\bfseries 12} (2023) 020} [\href{https://arxiv.org/abs/2306.01837}{{\ttfamily 2306.01837}}].

\bibitem{Gomez:2023tkr}
C.~Gomez, \emph{{Traces and Time: a de Sitter Black Hole correspondence}},  \href{https://arxiv.org/abs/2307.01841}{{\ttfamily 2307.01841}}.

\bibitem{Aguilar-Gutierrez:2023odp}
S.E.~Aguilar-Gutierrez, E.~Bahiru and R.~Esp\'\i{}ndola, \emph{{The centaur-algebra of observables}},  \href{https://arxiv.org/abs/2307.04233}{{\ttfamily 2307.04233}}.

\bibitem{Ong:2020xwv}
Y.C.~Ong, \emph{{Space\textendash{}time singularities and cosmic censorship conjecture: A Review with some thoughts}}, \href{https://doi.org/10.1142/S0217751X20300070}{\emph{Int. J. Mod. Phys. A} {\bfseries 35} (2020) 14} [\href{https://arxiv.org/abs/2005.07032}{{\ttfamily 2005.07032}}].

\bibitem{Nariai}
H.~{Nariai}, \emph{{On a New Cosmological Solution of Einstein's Field Equations of Gravitation}}, \href{https://doi.org/10.1023/A:1026602724948}{\emph{General Relativity and Gravitation} {\bfseries 31} (1999) 963}.

\bibitem{Chandrasekaran:2022eqq}
V.~Chandrasekaran, G.~Penington and E.~Witten, \emph{{Large N algebras and generalized entropy}}, \href{https://doi.org/10.1007/JHEP04(2023)009}{\emph{JHEP} {\bfseries 04} (2023) 009} [\href{https://arxiv.org/abs/2209.10454}{{\ttfamily 2209.10454}}].

\bibitem{Svesko:2022txo}
A.~Svesko, E.~Verheijden, E.P.~Verlinde and M.R.~Visser, \emph{{Quasi-local energy and microcanonical entropy in two-dimensional nearly de Sitter gravity}}, \href{https://doi.org/10.1007/JHEP08(2022)075}{\emph{JHEP} {\bfseries 08} (2022) 075} [\href{https://arxiv.org/abs/2203.00700}{{\ttfamily 2203.00700}}].

\bibitem{Akers:2019lzs}
C.~Akers, N.~Engelhardt, G.~Penington and M.~Usatyuk, \emph{{Quantum Maximin Surfaces}}, \href{https://doi.org/10.1007/JHEP08(2020)140}{\emph{JHEP} {\bfseries 08} (2020) 140} [\href{https://arxiv.org/abs/1912.02799}{{\ttfamily 1912.02799}}].

\bibitem{Freivogel:2018gxj}
B.~Freivogel and D.~Krommydas, \emph{{The Smeared Null Energy Condition}}, \href{https://doi.org/10.1007/JHEP12(2018)067}{\emph{JHEP} {\bfseries 12} (2018) 067} [\href{https://arxiv.org/abs/1807.03808}{{\ttfamily 1807.03808}}].

\bibitem{Aguilar-Gutierrez:2023tic}
S.E.~Aguilar-Gutierrez, A.K.~Patra and J.F.~Pedraza, \emph{{Entangled universes in dS wedge holography}},  \href{https://arxiv.org/abs/2308.05666}{{\ttfamily 2308.05666}}.

\bibitem{Aguilar-Gutierrez:2023zoi}
S.E.~Aguilar-Gutierrez and F.~Landgren, \emph{{A multiverse model in dS wedge holography}},  \href{https://arxiv.org/abs/2311.02074}{{\ttfamily 2311.02074}}.

\bibitem{Emparan:2022ijy}
R.~Emparan, J.F.~Pedraza, A.~Svesko, M.~Toma\v{s}evi\'c and M.R.~Visser, \emph{{Black holes in dS$_{3}$}}, \href{https://doi.org/10.1007/JHEP11(2022)073}{\emph{JHEP} {\bfseries 11} (2022) 073} [\href{https://arxiv.org/abs/2207.03302}{{\ttfamily 2207.03302}}].

\bibitem{Panella:2023lsi}
E.~Panella and A.~Svesko, \emph{{Quantum Kerr-de Sitter black holes in three dimensions}}, \href{https://doi.org/10.1007/JHEP06(2023)127}{\emph{JHEP} {\bfseries 06} (2023) 127} [\href{https://arxiv.org/abs/2303.08845}{{\ttfamily 2303.08845}}].

\bibitem{Hollowood:2021nlo}
T.J.~Hollowood, S.P.~Kumar, A.~Legramandi and N.~Talwar, \emph{{Islands in the stream of Hawking radiation}}, \href{https://doi.org/10.1007/JHEP11(2021)067}{\emph{JHEP} {\bfseries 11} (2021) 067} [\href{https://arxiv.org/abs/2104.00052}{{\ttfamily 2104.00052}}].

\bibitem{Kudler-Flam:2023qfl}
J.~Kudler-Flam, S.~Leutheusser and G.~Satishchandran, \emph{{Generalized Black Hole Entropy is von Neumann Entropy}},  \href{https://arxiv.org/abs/2309.15897}{{\ttfamily 2309.15897}}.

\bibitem{Faulkner:2024gst}
T.~Faulkner and A.J.~Speranza, \emph{{Gravitational algebras and the generalized second law}},  \href{https://arxiv.org/abs/2405.00847}{{\ttfamily 2405.00847}}.

\bibitem{vanderHeijden:2024tdk}
J.~van~der Heijden and E.~Verlinde, \emph{{An Operator Algebraic Approach To Black Hole Information}},  \href{https://arxiv.org/abs/2408.00071}{{\ttfamily 2408.00071}}.

\bibitem{Susskind:2021omt}
L.~Susskind, \emph{{De Sitter Holography: Fluctuations, Anomalous Symmetry, and Wormholes}}, \href{https://doi.org/10.3390/universe7120464}{\emph{Universe} {\bfseries 7} (2021) 464} [\href{https://arxiv.org/abs/2106.03964}{{\ttfamily 2106.03964}}].

\bibitem{Susskind:2022bia}
L.~Susskind, \emph{{De Sitter Space, Double-Scaled SYK, and the Separation of Scales in the Semiclassical Limit}},  \href{https://arxiv.org/abs/2209.09999}{{\ttfamily 2209.09999}}.

\bibitem{Rahman:2022jsf}
A.A.~Rahman, \emph{{dS JT Gravity and Double-Scaled SYK}},  \href{https://arxiv.org/abs/2209.09997}{{\ttfamily 2209.09997}}.

\bibitem{Goel:2023svz}
A.~Goel, V.~Narovlansky and H.~Verlinde, \emph{{Semiclassical geometry in double-scaled SYK}},  \href{https://arxiv.org/abs/2301.05732}{{\ttfamily 2301.05732}}.

\bibitem{Shaghoulian:2021cef}
E.~Shaghoulian, \emph{{The central dogma and cosmological horizons}}, \href{https://doi.org/10.1007/JHEP01(2022)132}{\emph{JHEP} {\bfseries 01} (2022) 132} [\href{https://arxiv.org/abs/2110.13210}{{\ttfamily 2110.13210}}].

\bibitem{Susskind:2021dfc}
L.~Susskind, \emph{{Black Holes Hint Towards De Sitter-Matrix Theory}},  \href{https://arxiv.org/abs/2109.01322}{{\ttfamily 2109.01322}}.

\bibitem{Gao:2019nyj}
P.~Gao and D.L.~Jafferis, \emph{{A traversable wormhole teleportation protocol in the SYK model}}, \href{https://doi.org/10.1007/JHEP07(2021)097}{\emph{JHEP} {\bfseries 07} (2021) 097} [\href{https://arxiv.org/abs/1911.07416}{{\ttfamily 1911.07416}}].

\bibitem{Gao:2023gta}
P.~Gao, \emph{{Commuting SYK: a pseudo-holographic model}},  \href{https://arxiv.org/abs/2306.14988}{{\ttfamily 2306.14988}}.

\bibitem{Bintanja:2023vel}
S.~Bintanja, B.~Freivogel and A.~Rolph, \emph{{Tunneling to Holographic Traversable Wormholes}},  \href{https://arxiv.org/abs/2308.00871}{{\ttfamily 2308.00871}}.

\bibitem{Banihashemi:2022jys}
B.~Banihashemi and T.~Jacobson, \emph{{Thermodynamic ensembles with cosmological horizons}}, \href{https://doi.org/10.1007/JHEP07(2022)042}{\emph{JHEP} {\bfseries 07} (2022) 042} [\href{https://arxiv.org/abs/2204.05324}{{\ttfamily 2204.05324}}].

\bibitem{Banihashemi:2022htw}
B.~Banihashemi, T.~Jacobson, A.~Svesko and M.~Visser, \emph{{The minus sign in the first law of de Sitter horizons}}, \href{https://doi.org/10.1007/JHEP01(2023)054}{\emph{JHEP} {\bfseries 01} (2023) 054} [\href{https://arxiv.org/abs/2208.11706}{{\ttfamily 2208.11706}}].

\bibitem{Jacobson:2022gmo}
T.~Jacobson and M.R.~Visser, \emph{{Entropy of causal diamond ensembles}},  \href{https://arxiv.org/abs/2212.10608}{{\ttfamily 2212.10608}}.

\bibitem{Hayden:2007cs}
P.~Hayden and J.~Preskill, \emph{{Black holes as mirrors: Quantum information in random subsystems}}, \href{https://doi.org/10.1088/1126-6708/2007/09/120}{\emph{JHEP} {\bfseries 09} (2007) 120} [\href{https://arxiv.org/abs/0708.4025}{{\ttfamily 0708.4025}}].

\bibitem{Maldacena:2019cbz}
J.~Maldacena, G.J.~Turiaci and Z.~Yang, \emph{{Two dimensional Nearly de Sitter gravity}}, \href{https://doi.org/10.1007/JHEP01(2021)139}{\emph{JHEP} {\bfseries 01} (2021) 139} [\href{https://arxiv.org/abs/1904.01911}{{\ttfamily 1904.01911}}].

\bibitem{Shenker:2013pqa}
S.H.~Shenker and D.~Stanford, \emph{{Black holes and the butterfly effect}}, \href{https://doi.org/10.1007/JHEP03(2014)067}{\emph{JHEP} {\bfseries 03} (2014) 067} [\href{https://arxiv.org/abs/1306.0622}{{\ttfamily 1306.0622}}].

\end{thebibliography}\endgroup
\end{document}